\documentclass[]{emulateapj-rtx4}

\usepackage{natbib,color} 
\bibpunct{(}{)}{;}{a}{}{,}

\DeclareRobustCommand{\ion}[2]{%
\relax\ifmmode
\ifx\testbx\f@series
{\mathbf{#1\,\mathsc{#2}}}\else
{\mathrm{#1\,\mathsc{#2}}}\fi
\else\textup{#1\,{\mdseries\textsc{#2}}}%
\fi}

\shorttitle{3D view of the thermal structure in a super-penumbral canopy}
\shortauthors{Beck, C.; Choudhary, D.P.; Rezaei, R.}

\begin{document}
\title{A three-dimensional view of the thermal structure in a super-penumbral canopy}


\author{C. Beck}
\affil{National Solar Observatory (NSO), USA}

\author{D.P. Choudhary}
\affil{Department of Physics \& Astronomy, California State University, Northridge, USA}

\author{R. Rezaei}
\affil{Kiepenheuer Institut f\"ur Sonnenphysik (KIS), Germany}




\begin{abstract}
We investigate the thermal topology in a super-penumbral canopy by determining the three-dimensional (3D) thermal structure of an active region. We derive the temperature stratifications in the active region by an inversion of the \ion{Ca}{ii} IR line at 854.2\,nm, assuming local thermal equilibrium (LTE). We trace the 3D topology of individual features located in the super-penumbral canopy, mainly radially oriented fibrils. We find that about half of the fibrils form short, arched, low-lying loops in the temperature cube. These closed loops connect from bright grains that are either in or close to the penumbra to the photosphere a few Mms away from the sunspot. They reach less than 1\,Mm in height. The other half of the fibrils rise with distance from the sunspot until they leave the \ion{Ca}{ii} IR formation height. Many of the fibrils show a central dark core and two lateral brightenings as seen in line-core intensity images. The corresponding velocity image shows fibrils that are as wide as the fibrils seen in intensity without a lateral substructure. Additionally, we study one example of exceptional brightness in more detail. It belongs to a different class of structures without prominent mass flows and with a 3D topology formed by two parallel, closed loops connecting patches of opposite polarity. We present evidence that the inverse Evershed flow into the sunspot in the lower chromosphere is the consequence of siphon flows along short loops that connect photospheric foot points. The dark-cored structure of the chromospheric fibrils cannot have an convective origin because of their location above regular granulation in an optically thin atmosphere. The dark core most likely results from an opacity difference between the central axis and the lateral edges caused by the significant flow speed along the fibrils.
\end{abstract}
\section{Introduction}
Sunspots in the solar photosphere are structured into a dark core, the umbra, and a slightly brighter ring, the penumbra, whose intensity is still below that of the surrounding convective granulation \citep[see, e.g.,][]{solanki2003,thomas+weiss2012}. The penumbra is structured into bright and dark, roughly radially oriented filaments. Observations in chromospheric lines reveal an outward continuation of the penumbra with a similar filamentary structure,  which has been termed super-penumbra \citep{loughhead1974}. 

As early as 1910, \citet{evershed1910} noted a pronounced difference between the flow patterns in the photosphere and chromosphere in and around sunspots. In the photosphere, Evershed found a displacement of spectral lines that could be explained by a radial outflow that was nearly parallel to the surface \citep[later termed ``Evershed flow'';][]{evershed1909}, while in the chromosphere he found exactly the opposite flow direction \citep[similarly later termed ``inverse Evershed flow'';][]{evershed1910,stjohn1911,stjohn1911a}. Numerous subsequent investigations found that the flow patterns are structured into thin, elongated filaments both in the photosphere \citep[e.g.][and references therein]{hirzberger+kneer2001,tritschler+etal2004,ichimoto+etal2007} and in the chromosphere \citep[e.g.][]{maltby1975,dialetis+etal1985,tsiropoula2000,georgakilas+etal2003}. Above the chromosphere, the transition region to the corona was found to also exhibit the inverse Evershed effect \citep{nicolas+etal1981,alissandrakis+etal1988,dere+etal1990}. 

It is generally assumed that in the chromosphere both dark filaments as seen in the Balmer line of H$_\alpha$ \citep[e.g.][]{foukal1971a,zirin1972,zirin1974} and dark or bright fibrils seen in \ion{Ca}{ii} lines \citep[][]{pietarila+etal2009,beck+etal2010} trace magnetic field lines \citep{foukal1971,nakagawa+etal1971,nakagawa+etal1973,schad+etal2013}. This association obtained primarily by visual means, i.e.~chromospheric fibrils and filaments appear to originate or terminate in locations of photospheric magnetic flux and exhibit a spatial structure resembling the expected magnetic field configuration, was questioned early by \citet{frazier1972} and recently by \citet{delacruzrodriguez+etal2011a}. A major problem is to establish the true magnetic connectivity \citep{simon+zirker1974,kawakami+etal1989,schrijver+title2003,reardon+etal2011} because often only photospheric magnetic measurements and chromospheric intensity measurements are available. \citet{foukal1971a} suggested a classification of structures seen in the H$_\alpha$ line by size and established connectivity, but more often than not its application lacks the knowledge of connectivity.

Related to the magnetic connectivity is the question of the physical driver of both the regular and inverse Evershed effect. Magnetic field lines that connect sites of significantly different magnetic field strength can lead to directed mass flows from the weaker to the stronger foot point \citep[``siphon flows'';][]{meyer+schmidt1968}. Siphon flows have been invoked for both the upper solar atmosphere \citep{cargill+priest1980,noci1981,georgakilas+etal2003a,lagg+etal2007} and the photosphere \citep{thomas1988,guglielmino+zuccarello2011}, and also for the regular Evershed effect \citep{thomas+montesinos1993,montesinos+thomas1997,schlichenmaier+etal1998,schlichenmaier+etal1998a}. One peculiarity of fast siphon flows is the generation of a (standing) shock in the down-flow branch \citep{thomas+montesinos1991}. Observational evidence for both siphon flows and the development of shocks at their abrupt termination has been presented by, e.g., \citet{degenhardt+etal1993}, \citet{uitenbroek+etal2006}, \citet{beck+etal2010} and \citet{bethge+etal2012}.

Confirmation of siphon flows as a possible source of the inverse Evershed flow in the chromospheric super-penumbra requires combining information on the chromospheric flow pattern with (photospheric) magnetic information. In addition, knowing the thermal structure of the solar chromosphere can provide a direct way to initially trace the thermal connectivity of structures, which most likely is identical to the magnetic connectivity as in the case of the dark filaments seen in H$_\alpha$. In this study, we describe the derivation of a temperature cube using an inversion of spectra observed in the \ion{Ca}{ii} IR line at 854.2\,nm. We use the inversion to determine the three-dimensional (3D) thermal topology of specific features in an active region (AR), primarily located in the super-penumbral canopy of the sunspot in the observed AR.

The analysis of chromospheric spectra is more difficult than for photospheric data. One reason is that only a handful of chromospheric spectral lines are accessible from the ground, e.g., the Balmer lines such as H$_\alpha$, H$_\beta$, etc..., neutral Helium lines such as \ion{He}{i} at 1083\,nm, or the lines of singly-ionized Calcium (\ion{Ca}{ii} H and K in the blue and the \ion{Ca}{ii} triplet in the near-infrared). In addition, the assumption of local thermal equilibrium (LTE) partially breaks down with increasing height in the stratified solar atmosphere because of the exponential decrease in particle density which in turn leads to an insufficient number of collisions to maintain thermal equilibrium. As a result, inversions of chromospheric line spectra \citep[e.g.][]{liu+skumanich1974,teplitskaia+etal1978,socasnavarro+etal2000,sheminova+etal2005,pietarila+etal2007,teplitskaya+grigoryeva2009,delacruzrodriguez+etal2012,martinezgonzalez+etal2012,beck+etal2013} are less common than their counterparts for photospheric spectra \citep[e.g.][and references therein]{keller+etal1990,cobo+toroiniesta1992,frutiger+etal2000,socasnavarro+etal2001,bellot+etal2004,carroll+kopf2008,asensio+etal2008,judge+carlsson2010,beck2011}. 

\ion{Ca}{ii} H and K and \ion{Mg}{ii} h and k \citep[e.g.][and references therein]{ayres+linsky1976,morrison+linsky1978,uitenbroek1997,leenaarts+etal2013} provide information from the continuum-forming layers in the outermost line wing up to heights of 2\,Mm or more in the line core. Because of the low line-core intensity of chromospheric lines (about 5\,\% of the continuum intensity), it is hard to measure their profiles with high signal-to-noise (S/N) ratios in sunspots or to determine the corresponding polarization signal.

The three \ion{Ca}{ii} infrared (IR) lines are interlocked with the H and K lines \citep[cf.][]{rutten+uitenbroek1991}, but have a smaller line-center opacity than the latter leading to an enhanced line-core intensity \citep{linsky+avrett1970}. For that reason, the \ion{Ca}{ii} IR lines are usually preferred in high-resolution observations with two-dimensional (2D) spectro(polari)meters such as the Interferometric BIdimensional Spectrometer \citep[IBIS;][]{cavallini2006,reardon+cavallini2008}, the CRisp Imaging Spectro-Polarimeter \citep[CRISP;][]{scharmer+etal2008} or the G\"ottingen/GREGOR Fabry-P{\'e}rot Interferometer \citep[GFPI;][]{puschmann+etal2006,puschmann+etal2012b}.
\begin{figure*}
\hspace*{11.4cm}\resizebox{4.2cm}{!}{\hspace*{.25cm}\includegraphics{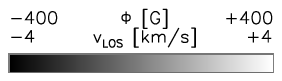}}\\\vspace*{-.2cm}\\
\resizebox{16.cm}{!}{\hspace*{1.cm}\includegraphics{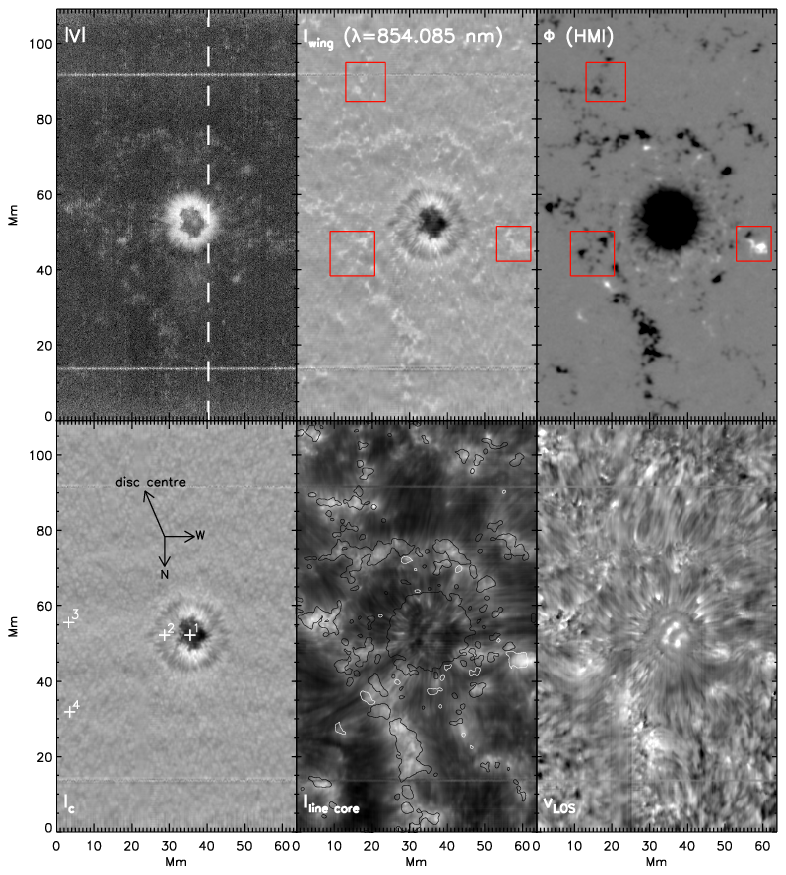}}$ $\\$ $\\$ $\\
\caption{Overview of the FOV. Bottom row, left to right: continuum intensity $I_c$, line-core intensity of Ca 854.2\,nm and line-core velocity of Ca 854.2\,nm. Top row, left to right: unsigned wavelength-integrated Stokes $V$ signal of Ca 854.2\,nm, line-wing intensity and pseudo-scan over the HMI magnetograms (see text). The white crosses in $I_c$ denote the locations of the profiles shown in Fig.~\ref{fig4}. The white/black contour lines in the line-core image mark magnetic fluxes of $\pm$50\,G. The vertical dashed line in the $|V|$-map indicates the location of the slit spectra shown in Fig.~\ref{fig2}. The red rectangles in the upper row outline distinct spatial features to facilitate a visual cross-check of the alignment. \label{fig1}}
\end{figure*}
\begin{figure*}
\centerline{\resizebox{6.5cm}{!}{\includegraphics{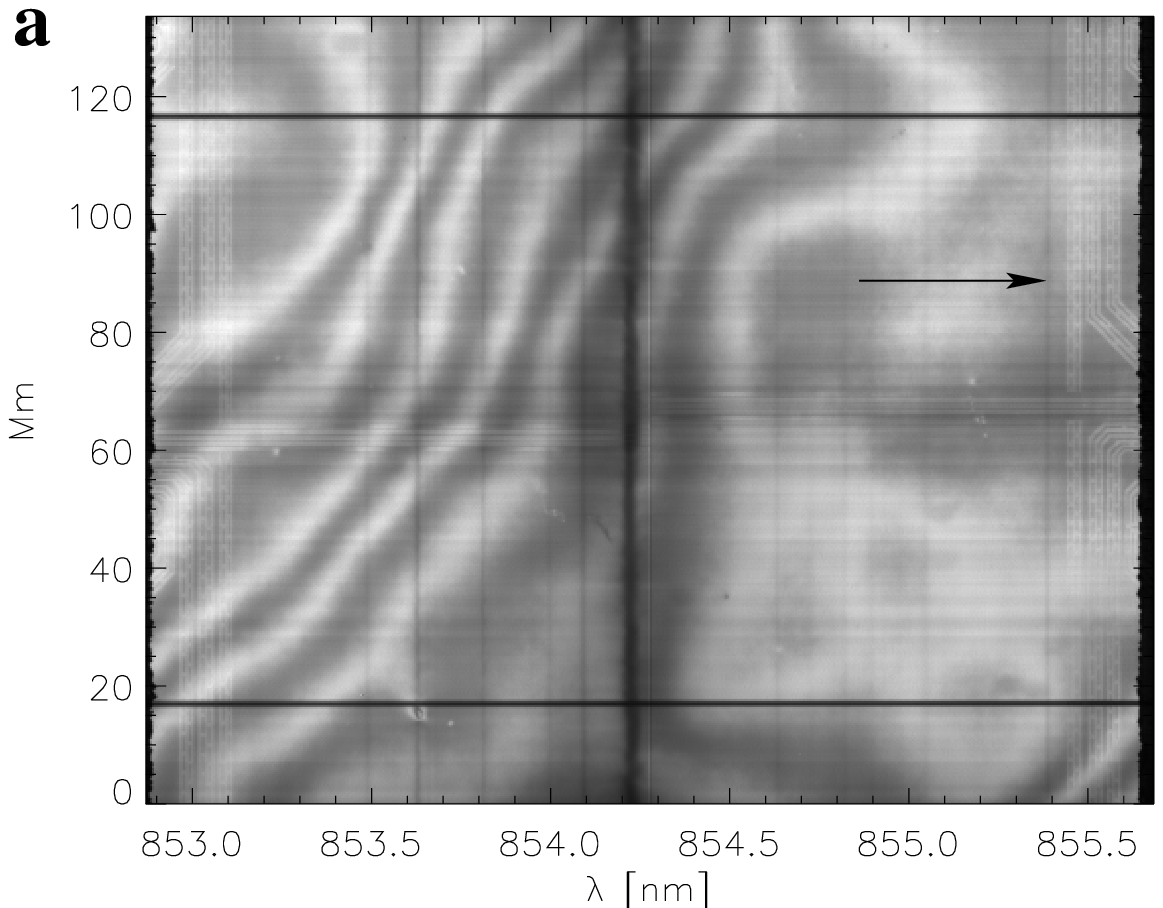}}\hspace*{.5cm}\resizebox{10.cm}{!}{\includegraphics{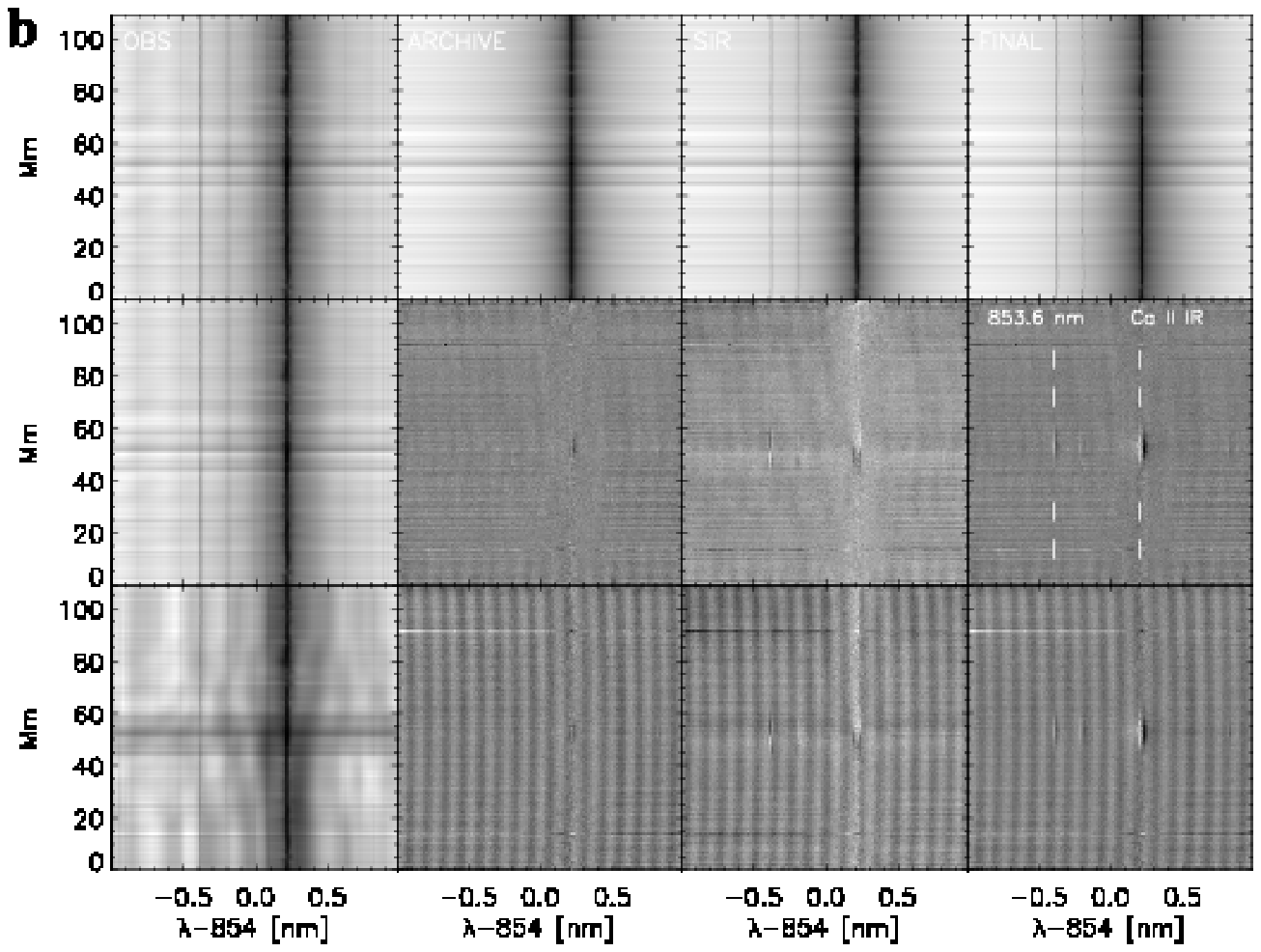}}}$ $\\
\caption{Panel a: raw spectrum of the Ca 854.2\,nm line. The black arrow points towards the circuit pattern imprinted on the spectra. Panel b: fringe filtering and inversion steps. Bottom row: flat-fielded spectra of Stokes $IQUV$ (left to right) from the scan step marked in Fig.~\ref{fig1}. Middle row: Stokes $IQUV$ after fringe filtering and extended flat fielding. The vertical dashed white lines indicate the line identification. Top row: observed spectra, best-fit LTE archive spectra (w/o blends and Doppler shifts), best-fit SIR inversion spectra and final best-fit spectra (LTE archive inversion with SIR LOS velocities).\label{fig2}}$ $\\
\end{figure*}

Here, we use spectra of the \ion{Ca}{ii} IR line at 854.2\,nm  to determine a 3D view of the solar chromosphere from an inversion of individual spectra collected over a 2D area on the solar surface. Section \ref{secobs} describes our observations. The data reduction is explained in detail in Sect.~\ref{secred}, while Sect.~\ref{secana} describes the data analysis. Our results are presented in Sect.~\ref{secres}, summarized in Sect.~\ref{secsumm} and discussed in Sect.~\ref{secdiss}. Section \ref{secconc} provides our conclusions. Appendix \ref{perform_fit} quantifies the quality of the inversion approach, while Appendix \ref{3dnetwork} shows an example of the 3D structure of a network element.
\section{Observations}\label{secobs}
On 29/07/2013 from 17:48 until 18:08 UT, we observed the active region (AR) NOAA 11801, comprising a fairly round and symmetric sunspot and some plage area at a heliocentric angle of about 20 degree, with the SPectropolarimeter for Infrared and Optical Regions \citep[SPINOR;][]{socasnavarro+etal2006} at the Dunn Solar Telescope (DST). We scanned the AR with 400 steps of 0\farcs22 ($\equiv$ the slit width), recording the full Stokes vector in the chromospheric \ion{Ca}{ii} IR line at 854.2\,nm (hence forth Ca 854.2\,nm). The exposure time for each image of the eight modulation states recorded was about 33\,ms. A set of eight measurements of the Stokes vector was added up to improve the S/N ratio. The spectral dispersion was 5.55 pm per pixel and the spatial sampling along the slit was 0\farcs36, respectively. The slit was oriented along solar North-South.

A second photospheric channel in the \ion{Fe}{i} lines at 1565\,nm was only added two days later, therefore we used line-of-sight (LOS) magnetograms from the Helioseismic and Magnetic Imager \citep[HMI;][]{scherrer+etal2012} on-board the Solar Dynamics Observatory \cite[SDO;][]{pesnell+etal2012} for photospheric magnetic context information. We created a ``pseudo-scan'' by stepping an artificial slit across a series of spatially aligned HMI magnetograms, using the pointing, step width and spatial sampling of the SPINOR observations \citep[cf.][Appendix B]{beck+etal2007}. The aligned HMI magnetogram closest in time to the SPINOR scan step was used to retrieve the corresponding slice of the pseudo-scan. In addition, we run an inversion of the HMI spectra taken at 18:00 UT using the Very Fast Inversion of the Stokes Vector code \citep[VFISV; cf.][]{borrero+etal2007,borrero+etal2011}. We inverted a region of 1k$\times$1k pixels centred on NOAA 11801 to also obtain an estimate of the LOS field inclination. \citet{kiess+etal2014} show an example of results obtained with the same settings for VFISV as used here.

Figure \ref{fig1} provides maps of the observed field of view (FOV) in line parameters derived from the Ca 854.2\,nm spectra and the HMI magnetogram pseudo-scan. The seeing was good to excellent during the observation. The spatial resolution of the data is clearly below 1$^{\prime\prime}$ (e.g.~the smallest discernible structures in the Ca line-core or line-wing image in Fig.~\ref{fig1}). The LOS velocity measured in the Ca 854.2\,nm line (lower right panel in Fig.~\ref{fig1}) shows large velocities of opposite sign in the umbra, as well as several roughly radially oriented fibrils in the super-penumbral canopy whose red-shift/down-flow velocity increases towards the sunspot. These fibrils have counterparts in the line-core intensity (lower middle panel) that usually show an additional sub-structure of a dark core flanked by two lateral brightenings at the location of the velocity fibrils (cf.~the magnifications in Fig.~\ref{fig7} below). The unsigned, wavelength-integrated Stokes $V$ signal observed in Ca 854.2\,nm in the upper left panel shows significant polarization signal inside the sunspot, but only faint signal in the surrounding network. This a consequence of the comparatively short integration time used in the observations (about 2.1 seconds per scan step). The appearance of the network in the $|V|$-map and in the line-wing intensity (upper middle panel) match well with the sites of magnetic flux in the HMI pseudo-scan (upper right panel in Fig.~\ref{fig1}), confirming the good alignment of the data.
\section{Data reduction}\label{secred}
Before analysing the data, it was necessary to apply a few additional steps not normally required in spectropolarimetric data reduction. 
\subsection{Standard flat field and polarimetric calibration}
The first step in the data reduction was a standard approach to flat fielding. A gain table was calculated from flat field data taken with a defocused telescope making a random walk during the exposures. We used a method similar to that described in \citet{beck+etal2005b}. The method uses several average
profiles from overlapping subdivisions along the slit to remove the spectral
line information and to obtain a correction for local intensity
inhomogeneities. The drawback of this approach is that features that extend the length of the slit in the spatial dimension, such as interference fringes, or whose extent is larger than the size of the subdivision from which the average
profile is calculated, are not removed by the gain correction. In the Ca 854.2\,nm wavelength range, the SARNOFF camera used unfortunately shows
strong, large-scale patterns that result from interference fringes and  a ghost image most likely of photons reflected by the read-out electronics (cf.~panel (a) of Fig.~\ref{fig2}). In Stokes $Q, U$ and $V$, only interference fringes were present (right three panels of bottom row in panel (b) of Fig.~\ref{fig2}). The Stokes parameters were obtained by the successive application of the inverse of the polarimeter response function \citep[e.g.][]{skumanich+etal1997,beck+etal2005b,beck+etal2010,socasnavarro+etal2011} and the telescope matrix \citep[e.g][]{capitani+etal1989,beck+etal2005a,socasnavarro+etal2011}.
\begin{figure}
\resizebox{8.4cm}{!}{\hspace*{.75cm}\includegraphics{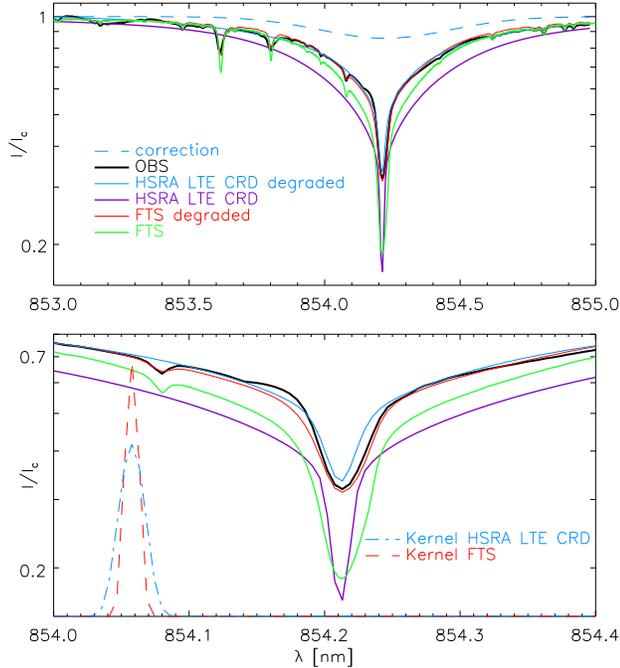}}$ $\\
\caption{Average observed and reference profiles with the full spectral range
  of the LTE archive (top panel) and restricted to the line-core region of
  Ca 854.2\,nm (bottom panel). Thick black: average observed
  profile. Thick green/thin red: original and degraded FTS atlas
  spectrum. Thick purple/thin blue: original and degraded LTE modified HSRA
  spectrum. Blue dashed: additional correction for the LTE HRSA profile. The red-dashed and blue dash-dotted lines at the left-hand side in the bottom
  panel show the spectral PSFs for degrading FTS and LTE HSRA profile to the observations.\label{fig3}}
\end{figure}
\begin{figure*}
\begin{center}
\hspace*{1cm}\resizebox{6.cm}{!}{\includegraphics{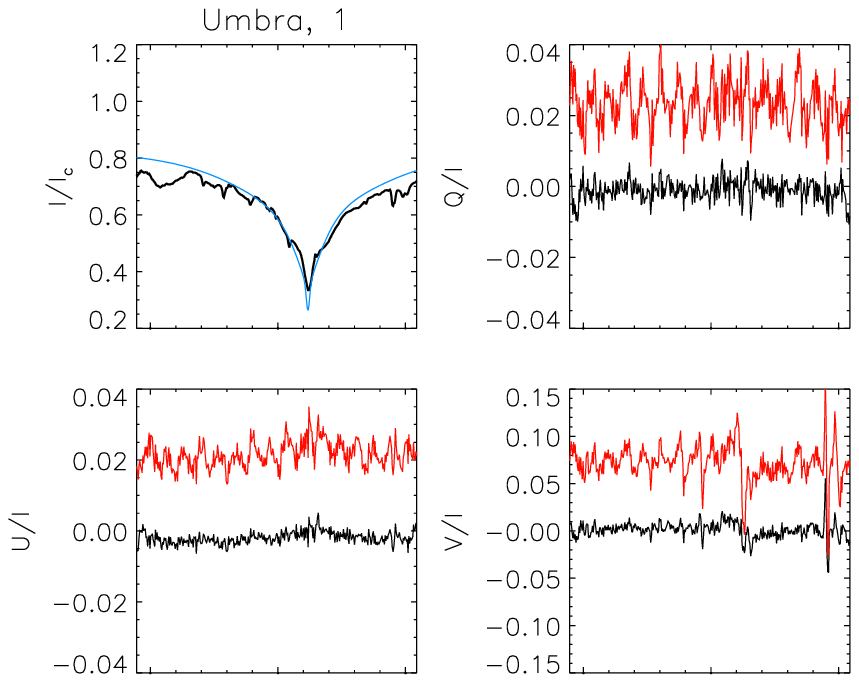}}\hspace*{.75cm}\resizebox{6.cm}{!}{\includegraphics{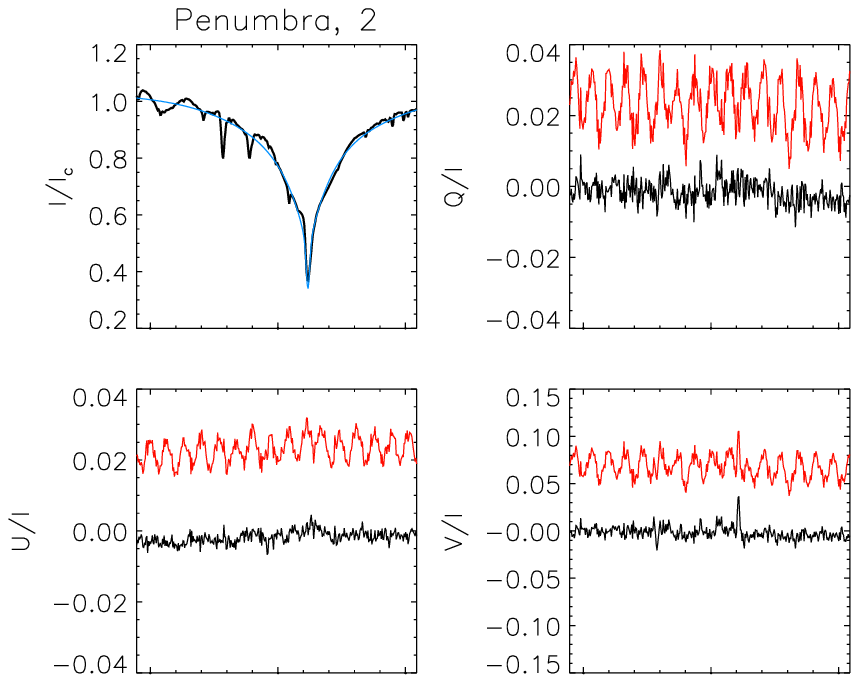}}\\$ $\\
\hspace*{1cm}\resizebox{6.cm}{!}{\includegraphics{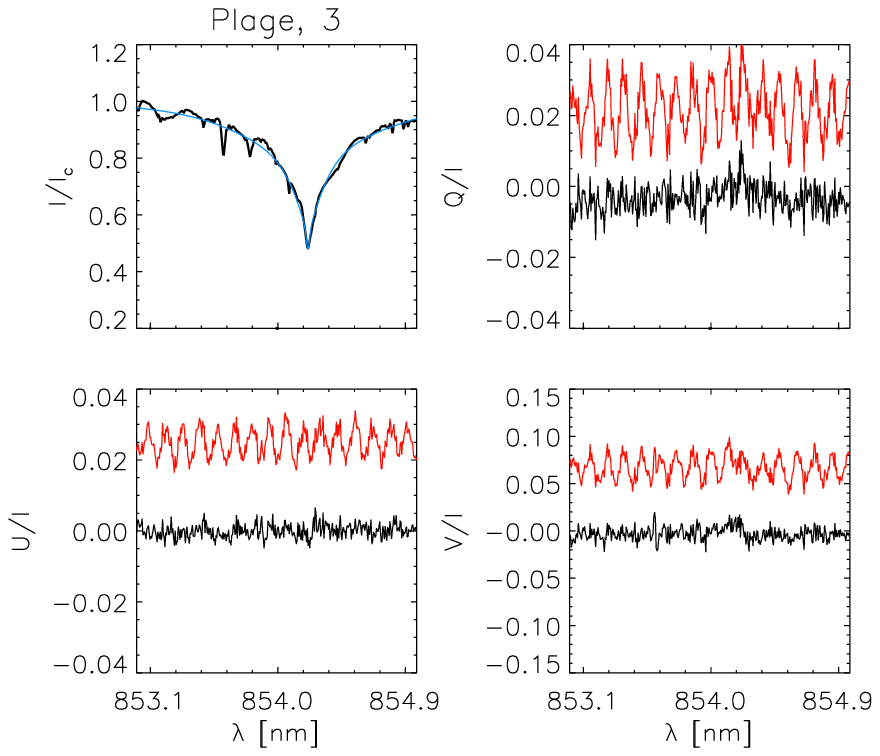}}\hspace*{.75cm}\resizebox{6cm}{!}{\includegraphics{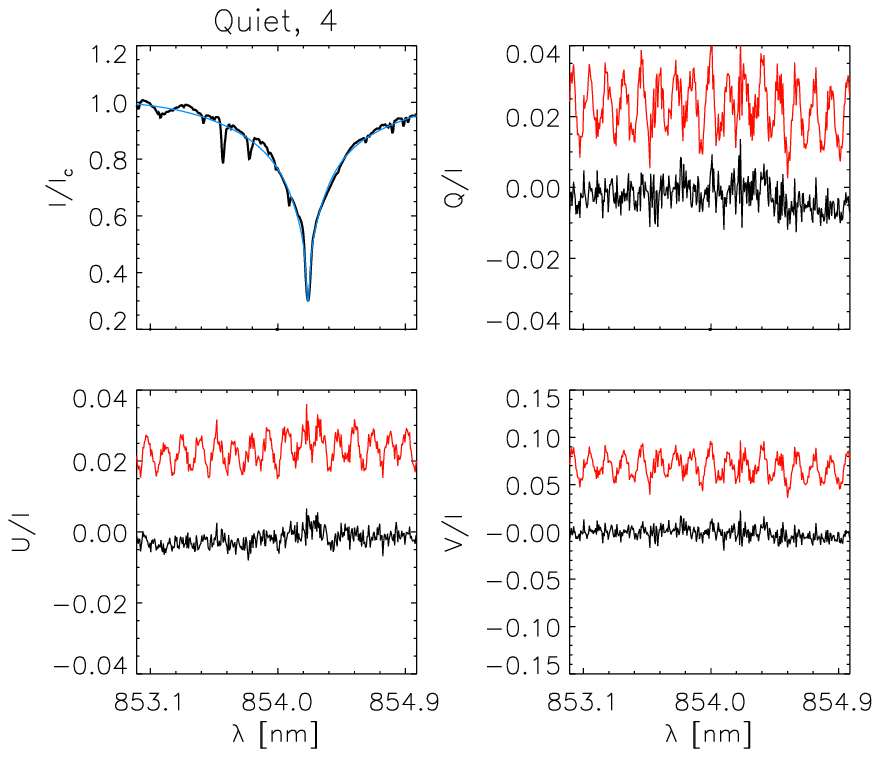}}$ $\\$ $\\
\end{center}
\caption{Example profiles at different locations in the FOV. Black: observed
  profiles after all corrections. Blue: best-fit LTE archive profile (Stokes
  $I$ only). Red: initial $QUV$ profiles without fringe correction, offset by
  0.025 in $QU$ and 0.075 in $V$ for better visibility. Top row: umbral and penumbral profiles. Bottom row: plage and quiet Sun profiles. \label{fig4}}
\end{figure*}
\subsection{Extended flat fielding and fringe correction}\label{flatfield}
To remove the residual large-scale patterns in Stokes $I$, we used the Fourier Transform Spectrometer \citep[FTS;][]{kurucz+etal1984,neckel1999} atlas spectrum  as reference. We first matched the average observed spectrum in the quiet Sun (QS) to the corresponding section of the FTS atlas in dispersion and observed wavelength range. We then created a pseudo-slit spectrum from the FTS profile with the same dimensions as an individual observed slit spectrum by expanding the FTS profile along the slit. Dividing the observed spectrum by the FTS pseudo-slit spectrum yielded a correction for the large-scale features in the spatial dimension and for any wavelength trends caused by, e.g., the order-sorting interference filter. This correction was determined for every scan step and applied to the Stokes $I$ spectra only (cf.~leftmost panel in the middle row in panel (b) of Fig.~\ref{fig2}). It unfortunately left some residuals at specific wavelengths on the scan steps, where the slit crossed the umbra. The enhanced brightness in the penumbra near the umbral boundary from $x \sim 32$ to $38$\,Mm in the line-wing intensity in Fig.~\ref{fig1} is such a residual that is not of solar origin.

For the polarization components Stokes $Q, U$ and $V$, we only applied a
conservative Fourier filter to the slit-spectra by removing the Fourier frequencies corresponding to approximately the full slit length (low spatial frequencies) in a range of spectral frequencies typical of the interference fringes. The approach worked well for quiet Sun spectra, but partially failed in the sunspot's umbra where it left fringes of significant amplitude. For umbral profiles, we therefore applied a second Fourier filter on individual $QUV$ profiles, removing the characteristic frequencies of the fringes that remained after the first correction on the full slit spectrum (cf.~the middle row in panel (b) of Fig.~\ref{fig2} and Fig.~\ref{fig4}). A more complex approach as discussed in \citet{casini+etal2012} seemed unnecessary. The root-mean-squared (rms) noise values in $QUV$ were $4 \times 10^{-3}, 2 \times 10^{-3}$, and $6 \times 10^{-3}$ of $I_c$ after the fringe filtering, respectively.
\subsection{Determination of spectral PSF and intensity normalization}
To invert the spectra, we matched the average observed spectrum in the QS with the FTS atlas and with a spectrum that was synthesized from a modified version of the Harvard Smithsonian Reference Atmosphere \citep[HSRA;][]{gingerich+etal1971} using the Stokes Inversion based on Response functions code \citep[SIR;][]{cobo+toroiniesta1992}. The SIR code assumes LTE and complete frequency re-distribution (CRD). In this modified version of the HSRA, the chromospheric temperature rise above $\log\,\tau = -4$ is replaced by a linear extrapolation of the temperature from lower layers (cf.~\citeauthor{beck+etal2013} \citeyear{beck+etal2013} (BE13); \citeauthor{beck+etal2013a} \citeyear{beck+etal2013a}). We used the approach of
\citet{allendeprieto+etal2004} and \citet{cabrerasolana+etal2007} that
consists of adding a wavelength-independent stray-light offset $\beta$ to the
reference spectrum, followed by a convolution with a Gaussian of
width $\sigma$. The corresponding values were $\beta = 17\,\%$ and $\sigma = 4.2\,$pm ($7\,\%$ and $8.3\,$pm) for matching th FTS (modified HSRA) with the average observed spectrum. After convolving the reference and synthetic spectra, an intensity normalization coefficient was obtained by matching the intensity in the line wing. 
\begin{figure*}
\centerline{\resizebox{17.6cm}{!}{\hspace*{1cm}\includegraphics{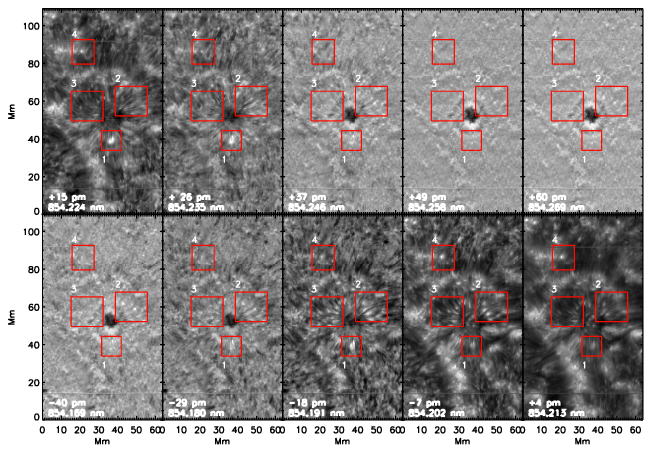}}}$ $\\
\caption{Full FOV in different wavelengths near the line core of Ca 854.2\,nm. The two large red squares mark regions 2 and 3 containing fibrils in the super-penumbral canopy. Region 1 marks a prominent brightening whose visibility is enhanced/reduced in the line wing/line core. Region 4 is centred on a magnetic flux concentration in the photospheric network.\label{fig5}}
\end{figure*}

Figure \ref{fig3} shows the average observed QS spectrum, the initial
profiles of FTS and synthetic spectra, and the final result after the convolution and intensity normalization. It was necessary to
apply two additional corrections to the synthetic HSRA spectrum to match the average observed profile, i.e.~a slight increase in the intensity in the line-core region and in the inner line wing (division with the blue dashed line in the top panel of Fig.~\ref{fig3}) and a multiplication of the full synthetic line profile by 1.03 after the convolution. These two corrections were not mutually independent and were simply defined ad-hoc. The final result was that the line shape of the average observed profile and the convolved and corrected synthetic profile matched acceptably (note that the panels in Fig.~\ref{fig3} are displayed on a logarithmic scale), which implies that the inversion of the average profile would yield a temperature stratification similar to the modified HSRA
model. 
\begin{figure}
\centerline{\hspace*{1.5cm}\resizebox{4.5cm}{!}{\includegraphics{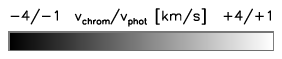}}}\vspace*{-.2cm}$ $\\
\resizebox{8.4cm}{!}{\hspace*{.3cm}\includegraphics{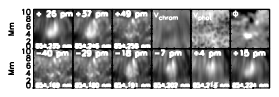}}\vspace*{.1cm}$ $\\
\resizebox{8.4cm}{!}{\hspace*{.3cm}\includegraphics{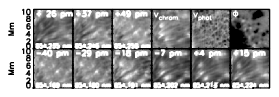}}\vspace*{.1cm}$ $\\
\resizebox{8.4cm}{!}{\hspace*{.3cm}\includegraphics{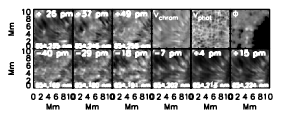}}$ $\\$ $\\$ $\\
\caption{Magnifications of the regions 1 to 3 (top to bottom) marked in Fig.~\ref{fig5} in different wavelengths near the Ca 854.2\,nm line core. Rightmost three panels in top row: chromospheric and photospheric LOS velocity, and magnetic flux clipped to $\pm$100\,G.\label{fig7}}
\end{figure}
\section{Data analysis}\label{secana}
\subsection{Inversion approach}
For the inversion of the Ca 854.2\,nm spectra, we used a two-step
approach. First, a best-fit spectrum and its corresponding temperature
stratification were obtained by comparing each individual spectrum to
an archive of pre-calculated LTE spectra. For Ca 854.2\,nm, it turned out that these archive spectra directly provided a good fit through the entire observed spectral range. This is contrary to \ion{Ca}{ii} H where line core and line wing are decoupled (BE13) because of the stronger intensity response to temperature variations at short wavelengths. On the other hand, we currently lack accurate information on the line blends of Ca 854.2\,nm, which prevented us from ascribing a reliable formation height to them, as done for instance for the blends of \ion{Ca}{ii} H in \citet{beck+etal2009}, \citet{felipe+etal2010}, \citet{bethge+etal2012} and BE13. We therefore run an inversion with the SIR code on the spectra as a second step to obtain a suitable velocity value, using the temperature stratification of the best-fit archive spectra as the initial model. 
\paragraph{Preparation of LTE archive}
To use the LTE archive for the inversion of the observed spectra, we first
synthesized the Ca 854.2\,nm line for the same set of temperature stratifications as used in BE13 for the inversion of \ion{Ca}{ii} H spectra. The synthesis used a spectral sampling of 4\,pm and did not include LOS velocities or line blends. A broadening by a macro-turbulent velocity of 1 km\,s$^{-1}$ was included in the synthesis. The synthetic spectra were spectrally re-sampled to match the dispersion of the observed spectra, and then treated with the same steps as determined above to match the synthetic spectrum from the modified HSRA with the average observed spectrum, i.e.~addition of $\beta$, convolution, application of correction around the line core and multiplication by 1.03. 
\paragraph{LTE archive fit}
To determine the archive spectrum that best fits an individual observed
spectrum, the least-square deviation between the observed profile and all of the archive spectra was calculated. Weights for specific wavelengths were chosen as in BE13 as the inverse of the average observed profile, which automatically gives the strongest weight to the Ca line core. The synthetic profile yielding the smallest least-square deviation delivers the corresponding LTE archive best-fit profile (e.g.~the second column in the top row of panel (b) of Fig.~\ref{fig2} or the blue lines in Stokes $I$ in Fig.~\ref{fig4}) with its corresponding temperature stratification. 
\paragraph{SIR inversion}
The LTE archive does not include different LOS velocities because it would become too large. Because we could not use the line blends to determine a velocity stratification for every profile, we run a SIR inversion on the observed spectra with the best-fit archive temperature stratification as the initial value. We used 11 nodes in temperature, but only one node in velocity, i.e.~a large number of degrees of freedom in temperature to match the line shape, but only one degree of freedom ($\equiv$ constant velocity with height) in
velocity. It turned out that any other choice with more degrees of freedom for
velocity or significantly fewer degrees of freedom in temperature persistently led to a random trade-off between temperature and velocity, with sometimes no convergence of the SIR fit at all \citep[see also][]{leenaarts+etal2014}. The final temperature stratifications determined by the SIR code usually were unrealistic with several reversals of the temperature gradient in optical depth. This results because the general line shape of the observed spectra, even after all corrections, still deviated from the FTS reference profile in the line wing (several ``bumps'' remained in the wing). We therefore discarded the temperature stratifications retrieved by the SIR code and only kept the LOS velocity which mainly reflects the Doppler shift of the strongest line in the spectrum, i.e.~Ca 854.2\,nm itself.
\paragraph{Final best-fit spectra}
The final-best fit spectra were synthesized by combining the temperature
stratifications from the best-fit archive spectra with the LOS velocities
retrieved by the SIR code in the second inversion step, and including the line
blends in the wing. An example is shown in the rightmost panel of the top row
of panel (b) of Fig.~\ref{fig2}. This synthesis provides no new quantitative information, but allows us to compare images in observed and final best-fit spectra at the same wavelengths. Figures \ref{fig5} and \ref{fig6} show the full FOV at a few wavelengths close to the line core of Ca 854.2\,nm in the observations and the final best-fit spectra, respectively. The inversion was able to match acceptably most large-scale features and isolated structures in the QS, but partly failed inside the umbra of the sunspot (cf.~also the upper left panel in Fig.~\ref{fig4}). The reason is that the low temperatures of the umbra with additional temperature reversals in the upper atmospheric layers were not included in the initial creation of the LTE archive which was constructed for inverting QS spectra. The maps of the intensity near the line core (854.208 to 854.244\,nm) in the best-fit spectra shown in Fig.~\ref{fig6} are somewhat coarser than the observed intensities in Fig.~\ref{fig5}. This results because the ``perfect'' match to the observed profile was not contained in the discrete, coarse archive, also outside of the umbra. The fit quality is quantified in appendix \ref{perform_fit}.
\section{Results}\label{secres}
\subsection{Definition of regions of interest}
Using the display of photospheric and chromospheric structure in
Fig.~\ref{fig5}, we defined four regions in the FOV for a
more detailed study. Region 1 in Fig.~\ref{fig5} marks a prominent brightening in both the blue and red line wings close to the line core, while in the line core itself this region is less prominent than most other places in the network. Regions 2 and 3 correspond to slender, fibrillar structures in the super-penumbra that originate (or terminate) in or close to the penumbra of the sunspot. Region 4 is centred on a magnetic flux concentration that is part of the photospheric magnetic network. It was kept in this analysis as a separate example for the capability of the inversion to derive the 3D thermal topology of structures in the solar photosphere and chromosphere and is discussed in Appendix \ref{3dnetwork}.

The regions 1 to 3 are shown in magnification in Fig.~\ref{fig7}. For regions 1 and 2, we find that many of the fibrils consist of a central dark core flanked
by two lateral brightenings at wavelengths close to the line core. The corresponding velocity fibrils are as wide as all parts of the respective intensity fibril together, i.e.~they cover both the dark core and the lateral brightenings seen in intensity. The fibrils in region 2 on the limb side of the sunspot either terminate in bright grains in the outer penumbra or in some cases just outside of the sunspot. The LOS velocity seen in Ca 854.2\,nm shows a clear increase towards the end of the fibrils located in the sunspot. The photospheric LOS velocity (upper last-but-one panel to the right in each row in Fig.~\ref{fig7}) derived from the Si line blend shows no clear indications of the velocity fibrils. The brightening at the line core in region 1 seems to belong to a different class of features. It is spatially more extended and has no clear counterpart in the chromospheric LOS velocity. 

In region 1, the LOS magnetogram exhibits a large patch with opposite polarity (white) from that of the sunspot (black). In region 2, the fibrils overlay faint patches of opposite polarity without terminating in them, while in region 3 the fibrils do not spatially align with the patches of opposite polarity.
\subsection{Animation of $T(\log \tau)$}
We created an animation (animation 1: anim\_temperature\_tau\_2d.avi\footnote{Available in the on-line section.}) that shows a scan in successive optical depths of temperature on the left and the pseudo-scan of the magnetogram at different threshold values on the right. While proceeding from the continuum-forming layers towards the chromosphere, the spatial connectivity of individual fibrils and the lateral expansion with height of almost all network elements is apparent in the temperature maps. In addition to the regions defined in Fig.~\ref{fig5}, a second feature (blue rectangle in the animation) with a similar appearance to that in region 1 can be seen. This second feature is less prominent in Fig.~\ref{fig5}, but shows the same general structure as region 1 in the animation. In particular, it shares with region 1 a location close to or above a magnetic patch with a polarity opposite to that of the sunspot.
\begin{figure}
\begin{centering}
\centerline{\hspace*{-1cm}\resizebox{3.7cm}{!}{\includegraphics{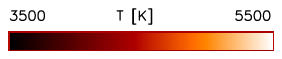}}}
\resizebox{7.5cm}{!}{\includegraphics{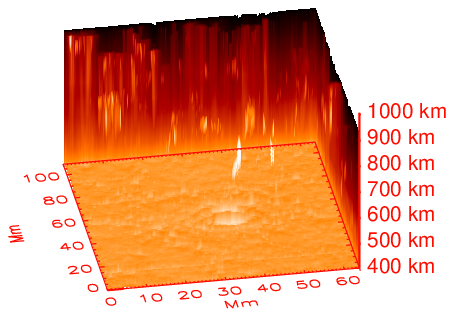}}\\
\resizebox{7.5cm}{!}{\includegraphics{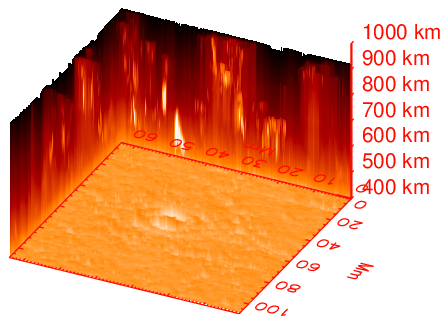}}\\
\resizebox{7.5cm}{!}{\includegraphics{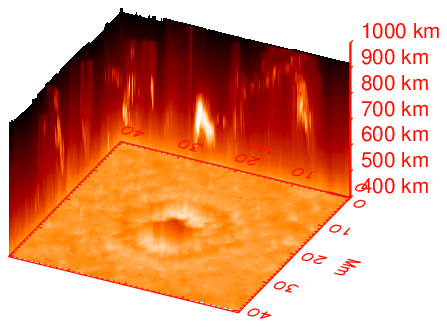}}\\
\end{centering}
\caption{3D rendering of the temperature in the full FOV from different
  viewing angles (top and middle panel) and magnification of the sunspot and its closer surroundings (bottom panel). Display range for temperatures is 3500 to 5500\,K (same for Figs.~\ref{fig9} and \ref{fig14} below).
\label{fig8}}
\end{figure}
\begin{figure*}
\begin{centering}
\resizebox{5.5cm}{!}{\includegraphics{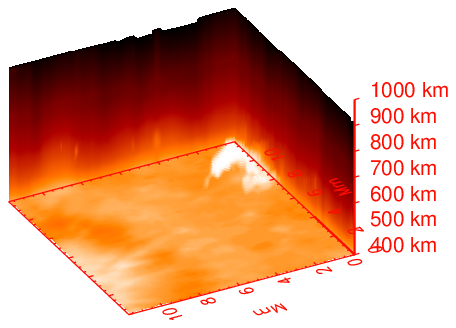}}\hspace*{.1cm}\resizebox{5.5cm}{!}{\includegraphics{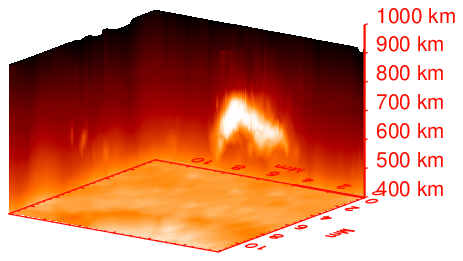}}\hspace*{.1cm}\resizebox{5.5cm}{!}{\includegraphics{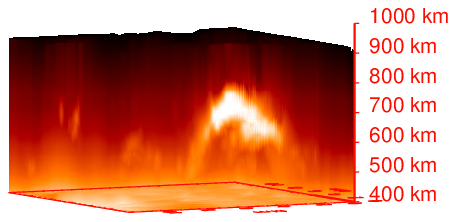}}\\
\end{centering}
\caption{3D rendering of the temperature in region 1 from different viewing angles. Left to right: viewing angle relative to the horizontal plane is 55/15/5 deg; the angle relative to the $z$-axis is 118, 145, and 118 degrees. See also animation 2 in the on-line section.\label{fig9}}
\end{figure*}
\begin{figure*}
\begin{centering}
\resizebox{5.5cm}{!}{\includegraphics{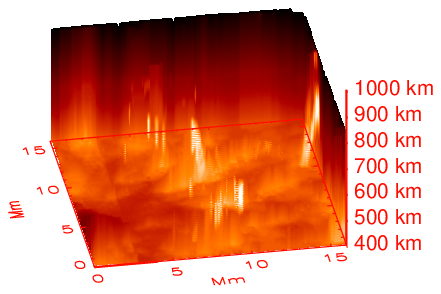}}
\hspace*{.1cm}\resizebox{5.5cm}{!}{\includegraphics{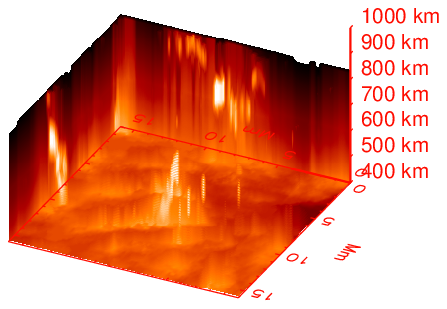}}
\hspace*{.1cm}\resizebox{5.5cm}{!}{\includegraphics{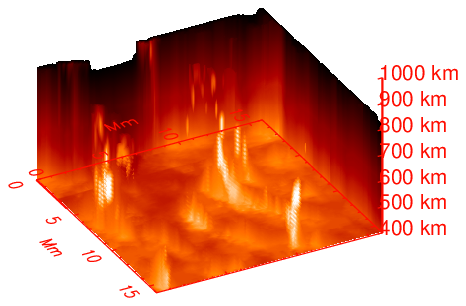}}\\
\resizebox{5.5cm}{!}{\includegraphics{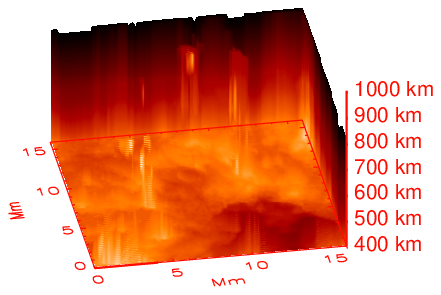}}
\hspace*{.1cm}\resizebox{5.5cm}{!}{\includegraphics{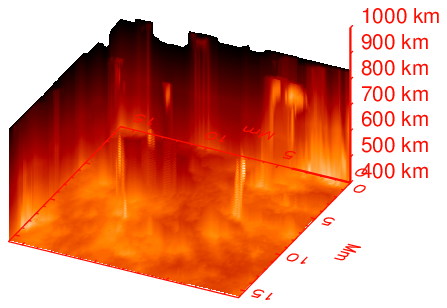}}
\hspace*{.1cm}\resizebox{5.5cm}{!}{\includegraphics{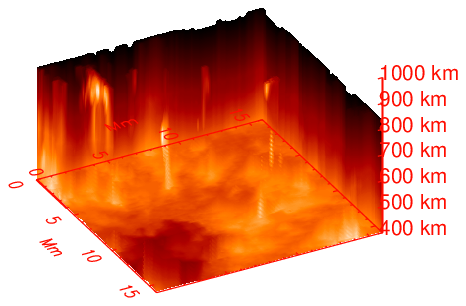}}\\
\end{centering}
\caption{3D rendering of temperature in regions 2 (top row) and 3 (bottom row) from different viewing angles. Viewing angle relative to the horizontal plane is 35 deg; the angle relative to the $z$-axis is 10, 154, and 298 degrees. See also animation 3 in the on-line section.\label{fig14}}
\end{figure*}
\subsection{3D rendering of temperature}
\paragraph{Full FOV and sunspot umbra} Figure \ref{fig8} shows a three-dimensional (3D) rendering of the temperature in the full FOV providing a general overview of its thermal structure, and a close-up of the region around the sunspot. The plots were generated from the 3D temperature cube\footnote{Available as IDL save file at\\ \url{ftp.nso.edu/outgoing/cbeck/temp\_cube/temp\_cube.sav}.} provided by the temperature stratifications on each pixel using the IDL function voxel\_proj. We used its option to set the intensity value of each ray to the maximum temperature value encountered along the ray. Because of the strong temperature gradient between photosphere and chromosphere, the bottom layers ($\log\,\tau > -1$) needed to be excluded. The scale along the $z$-axis is very rough since there was no explicit conversion from optical depth to geometrical height. The viewing angle relative to the horizontal plane or the $z$-axis can be freely chosen. 

The most outstanding features in Fig.~\ref{fig8} are a faint ring of slightly enhanced temperature just outside the outer penumbral boundary \citep[cf.][]{rast+etal2001,kitchatinov+ruediger2007,moradi+etal2010}, and an isolated brightening close to the sunspot that corresponds to the feature in region 1.
\paragraph{Region 1} Figure \ref{fig9} and the corresponding animation 2 (anim\_3d\_a.avi\footnote{Available in the on-line section.}) show 3D renderings of the temperature in region 1 from different viewing angles. The bright feature forms roughly an arched loop that starts in the photosphere close to the sunspot and bends down to the photosphere again within only a few Mm. From animation 2, we can see that the brightening actually consists of two loop structures, a shorter loop further from the sunspot and a longer one parallel to it, lying side by side in a roughly radial direction from the sunspot's centre. The connection down to the photosphere has a significantly lower temperature and a smaller diameter than the brightest part of the loop. The diameter is much smaller than the length of the structure.
\paragraph{Regions 2 and 3} The fibrils seen in regions 2 and 3 (Fig.~\ref{fig14}) show up as temperature ridges in a series of alternating increased and decreased temperatures. In region 2\footnote{Animation 3 anim\_3d\_fibrils\_1\_small.avi is available in the on-line section.}, there are several strong brightenings that correspond to the fibril heads near the sunspot, and in one case to a presumably unrelated network element (at $x,y \sim 15, 10$\,Mm in top row of Fig.~\ref{fig14}). The ridge structure is better visible on the limb side of the spot inside region 2 (top row of Fig.~\ref{fig14}). Also in region 3, the flow pattern of the fibrils is slightly less structured (cf.~Figs.~\ref{fig1} and \ref{fig7}).
\begin{figure}
\resizebox{8.cm}{!}{\hspace*{.75cm}\includegraphics{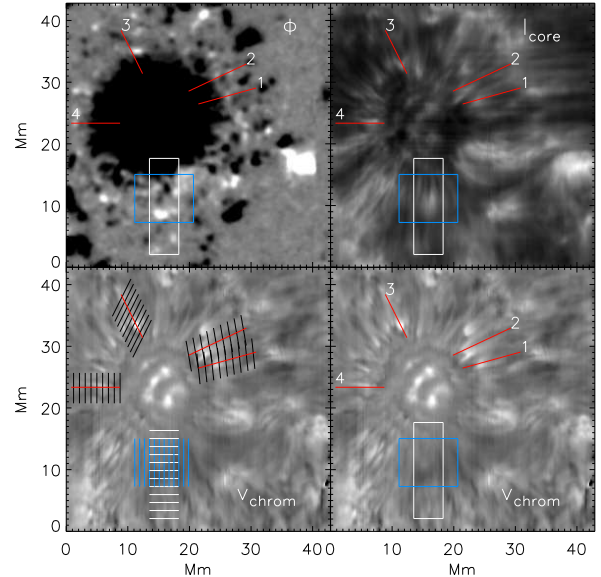}}$ $\\$ $\\
\caption{Overview of 2D spatial cuts through the temperature cube. Bottom row, background image: line-core velocity  of Ca 854.2\,nm. Top row: photospheric magnetic flux (left) and line-core intensity of Ca 854.2\,nm (right). Red lines mark the spines of fibrils, black lines every 7th cut perpendicular to the spines. The blue and white lines/rectangles mark the cuts along $y$ and $x$ in region 1, respectively. \label{cuts_over}}
\end{figure}
\begin{figure*}
\begin{centering}
\resizebox{16.1cm}{!}{\includegraphics{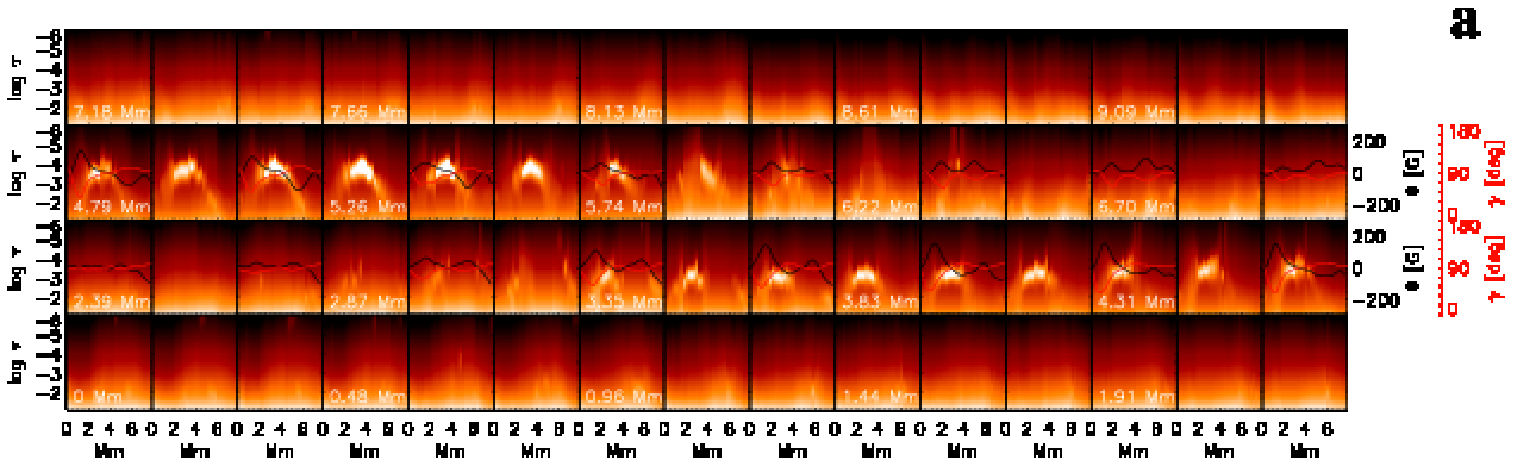}}\\
\resizebox{16.1cm}{!}{\includegraphics{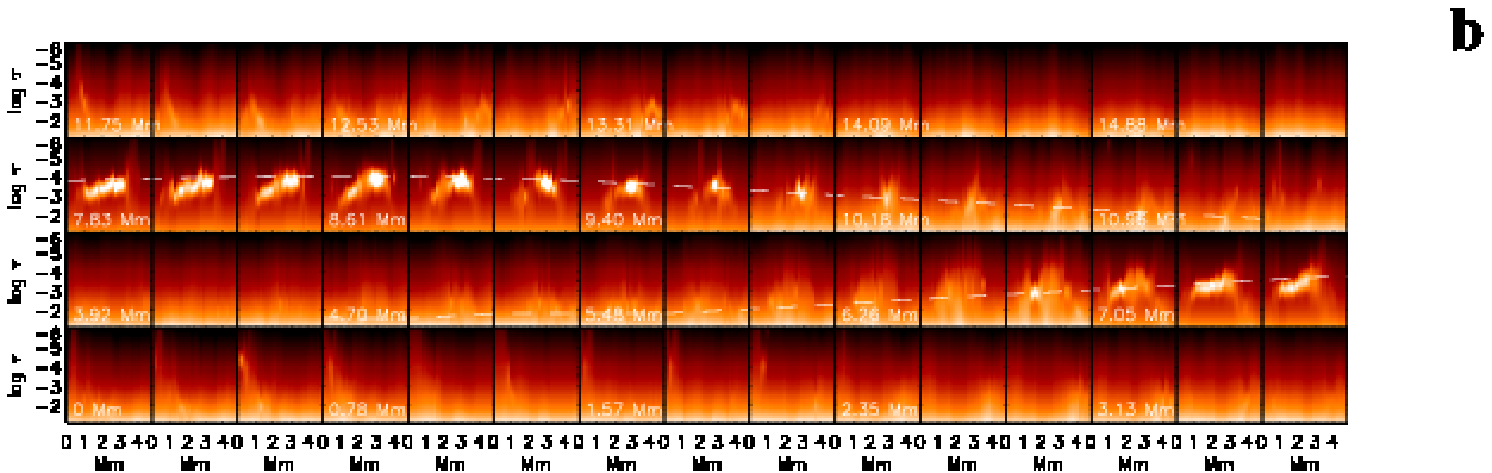}}\\
\hspace*{5cm}\begin{minipage}{1cm}
\resizebox{.75cm}{!}{\includegraphics[angle = 90]{zbar_temp.ps}}
\end{minipage}
\begin{minipage}{9cm}
\resizebox{8cm}{!}{\includegraphics{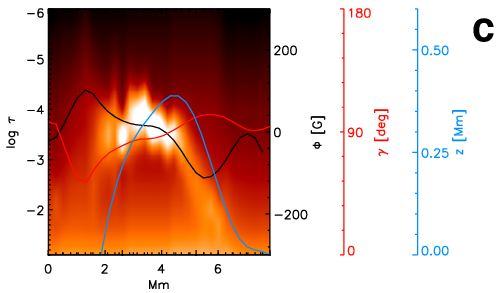}}
\end{minipage}
\end{centering}
\caption{Cuts through region 1 (cf.~Fig.~\ref{cuts_over}) along the $y$-axis (panel a) and along the $x$-axis (panel b). For the cuts in $y$ ($x$), the position in $x$ ($y$) increases from left to right in each row and from bottom to top between rows. The distances from the first position are given at the bottom of the panels for every third step.  The black and red lines in the middle rows of panel (a) give the LOS magnetic flux and the LOS inclination, respectively, along the cuts for every second sub-panel. The dashed white line in the middle rows of the panel (b) indicates the evolution of the height of the brightening in the atmosphere. Panel (c) shows a magnification of the third panel in the top-but-one row of panel (a). Here, the blue line additionally shows the integration of the LOS inclination along the cut assuming a vertical LOS and using the inner foot point as start point.\label{fig13}}
\end{figure*}
\subsection{Cuts through temperature cube}
To more accurately trace the spatial evolution of the topology, we laid cuts through the temperature cube to obtain a series of 2D horizontal-vertical temperature slices along the fibrils (cf.~Fig.~\ref{cuts_over}). For the fibrils in regions 2 and 3, we defined their spines by inclined straight lines that lie along the fibrils, and then took cuts perpendicular to the spines. Theses cuts trace the topology with increasing distance from the sunspot. For region 1, we made horizontal and vertical cuts along the $x$ and $y$ directions, because the structures are roughly aligned with the $y$-axis.
\paragraph{Region 1} Figure \ref{fig13} shows cuts through region 1 along the $y$-axis (panel a) and the $x$-axis (panel b). We stepped with the spatial sampling of the observations, i.e.~0\farcs22 in $x$ and 0\farcs36 in $y$, cutting out a slice of about 5 and 8\,Mm length, respectively. The cuts along the $y$-axis show clearly the loop shape of the feature, with its end points going down towards the photosphere on both sides of its apex. The overlaid cuts through the co-spatial and co-temporal magnetogram reveal that the loop seen in temperature connects two patches of opposite polarity that both lie outside of the sunspot (cf.~the upper border of the blue rectangle in Fig.~\ref{cuts_over}). The foot point closer to the sunspot (located towards the right in each sub-panel of panel (a) and in panel (c) in Fig.~\ref{fig13}) shares its polarity (negative), while the outer foot point is of positive polarity. 
\begin{figure*}
\begin{centering}
\hspace*{1cm}\resizebox{13cm}{!}{\includegraphics{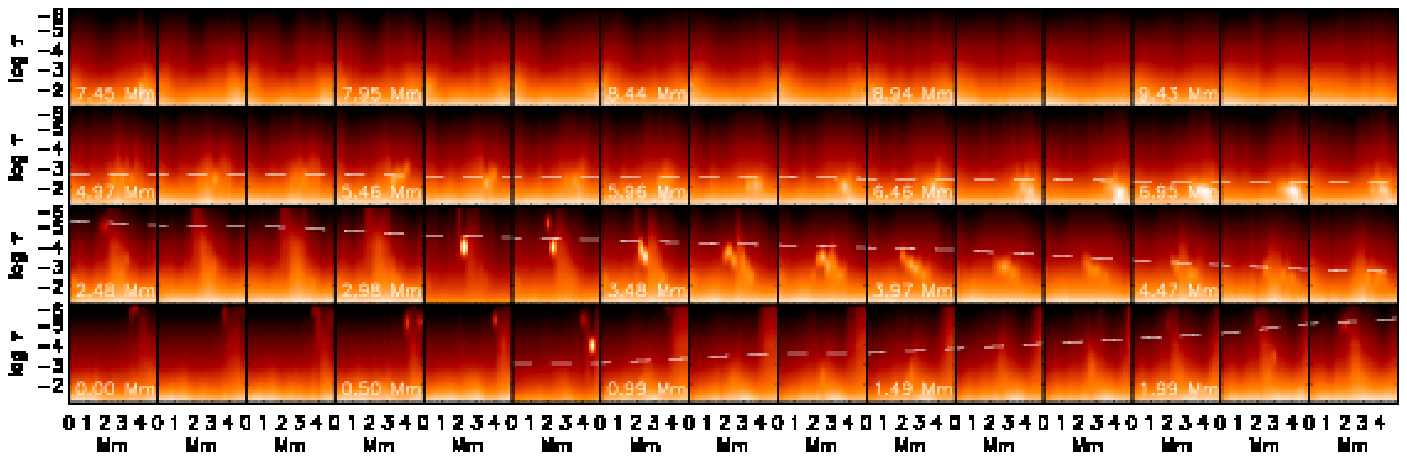}}\\
\hspace*{1cm}\resizebox{13cm}{!}{\includegraphics{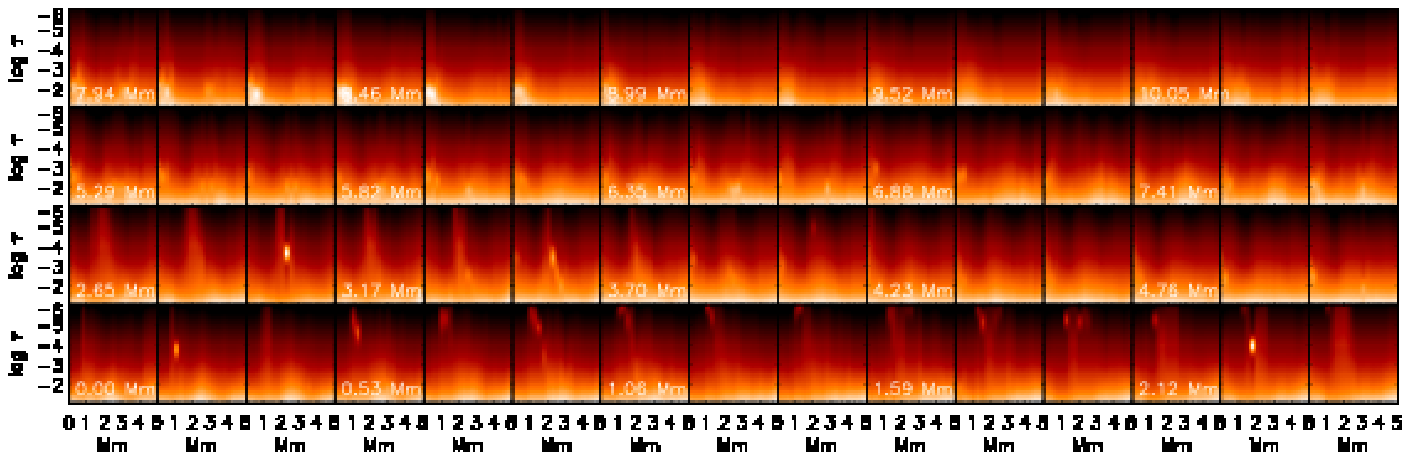}}\\
\hspace*{1cm}\resizebox{13cm}{!}{\includegraphics{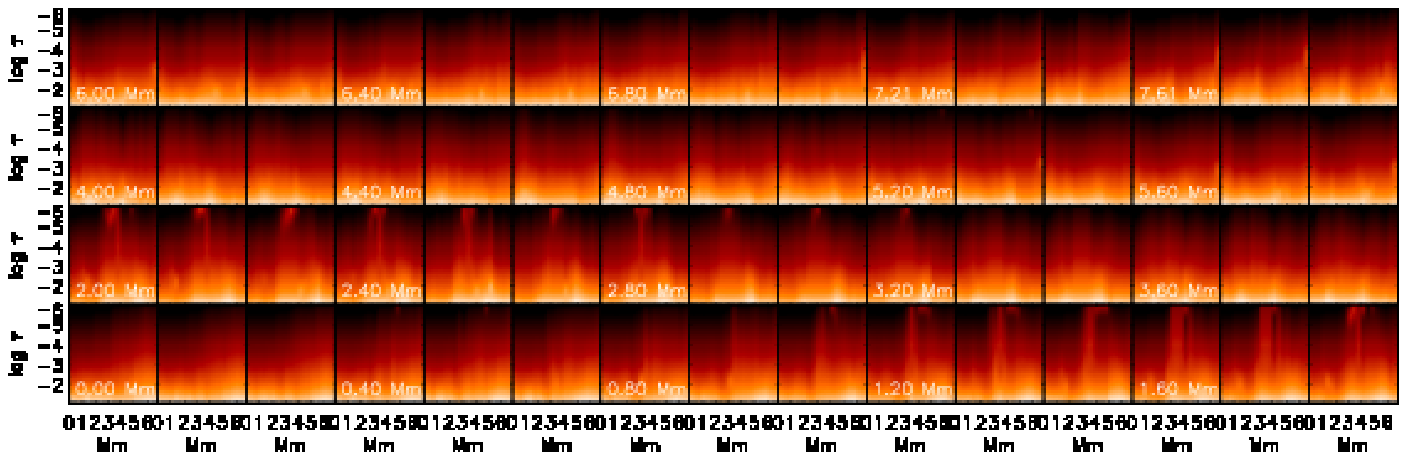}}\\
\hspace*{1cm}\resizebox{13cm}{!}{\includegraphics{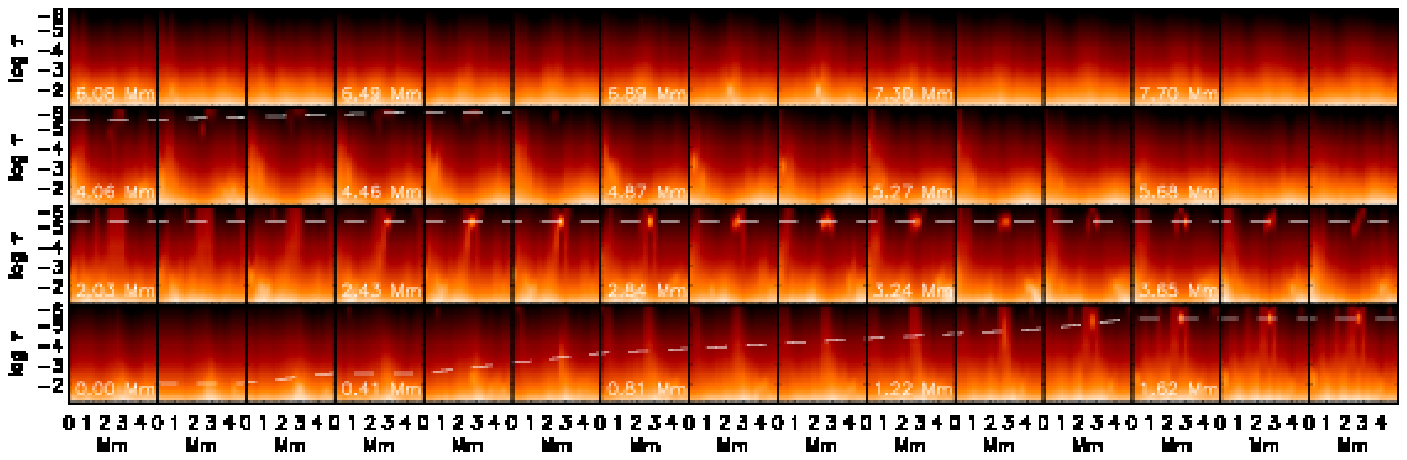}}\\
\end{centering}
\caption{$x-z$-slices of the temperature cube along the spines 1 to 4 (top to bottom) marked in Fig.~\ref{cuts_over}. Same layout as in Fig.~\ref{fig13}.\label{fig_cuts}}
\end{figure*}

The inclination of the photospheric magnetic field with respect to the LOS passes about 90$^\circ$ in the middle of loop. If we assume that the LOS is roughly perpendicular to the solar surface, we can integrate the LOS inclination along the cuts to obtain the height variation of the field lines \citep[cf.][]{solanki+etal2003a,beckthesis2006,beck2008}. The height variation of magnetic field lines derived by this approach follows the structure seen in temperature rather close (blue line in panel (c) of Fig.~\ref{fig13}). This implies that the loop should have had a significant polarization signal in the photospheric 617.3\,nm line used by HMI. Note that both the temperature and the integration of the LOS inclination of the magnetic field lines use LOS quantities, i.e.~both suffer the same projection effects from the inclined LOS with respect to the solar surface. The good agreement in the loop shape in temperature and integrated LOS field inclination, and especially the very similar apex height, offer an independent method to verify assigning formation heights to the optical depth layers of the temperature stratification \citep[cf., e.g.,][]{puschmann+etal2010,beck2011}.

In the cuts along the $x$-axis, the roughly roundish shape of the brighter and longer structure shows up (panel (b) of Fig.~\ref{fig13}). With increasing distance to the sunspot, the brightening becomes fuzzier and moves downward in height. The loop length is about 5 to 6\,Mm, with the apex at about $\log\,\tau = -4$ ($z \sim 0.5$\,Mm), where both values can be read off from, e.g., the magnification in panel (c) or panel (b) in Fig.~\ref{fig13} (start point at 5.48\,Mm and end point at 10.96\,Mm).

In total, we can confirm the magnetic connectivity with both foot points for the loops in region 1. We can also trace the complete thermal connectivity of the loop structure in the chromosphere and find that it is similar in shape and apex height to an integration of the photospheric LOS field inclination.
\paragraph{Regions 2 and 3} The cuts perpendicular to the fibrils for spines 1 to 4 are shown in Fig.~\ref{fig_cuts}. For spines 1 and 4, the spatial evolution of the topology is fairly apparent. Spine 1 (top panel of Fig.~\ref{fig_cuts}) displays a rising and then descending temperature enhancement as distance from the sunspot increases. Spine 4 (bottom panel of Fig.~\ref{fig_cuts}) shows a temperature enhancement whose height in the atmosphere increases with increasing distance to the sunspot until it leaves the formation height of Ca 854.2\,nm. For spines 2 and 3 the structure is more diffuse, but one could eventually trace a rise and descent of a temperature enhancement for spine 2 (second panel from the top) and a monotonically rising structure for spine 3 (second panel from the bottom). 

The central dark cores with lateral brightenings seen in the line-core intensity of Ca 854.2\,nm (e.g.~Fig.~\ref{fig7}) are also seen in the temperature stratifications in some cases. This is especially visible in the bottom panel of Fig.~\ref{fig_cuts}, as a central, roundish temperature enhancement high up in the atmosphere with a reduced temperature vertically below it, flanked by enhanced temperatures from $\log\,\tau = -5$ downward to the lower atmospheric layers. Given the limitations of the LTE approach, the corresponding profiles would be ideal candidates for a non-LTE inversion approach to determine the atmospheric properties at these locations more reliably. 
\section{Summary}\label{secsumm}
We have described the application of a rather simple and straightforward inversion procedure for spectra of the chromospheric \ion{Ca}{ii} IR line at 854.2\,nm based on the SIR code in synthesis mode. The approach retrieves LOS temperature stratifications under the assumption of local thermal equilibrium. We demonstrate that even within all the limitations related to the approach, its application to a chromospheric spectral line yields information on the complex topology of the solar chromosphere. We analyzed the spatial evolution of fibrils in the super-penumbral canopy and one exceptional strong brightening in wavelengths close to, but outside of the line core of Ca 854.2\,nm. All of the fibrils examined here either connect directly into the sunspot penumbra or close to it. We found that in our limited sample about half of the structures exhibit the shape of short loops of a few Mm extent that connect again down to the photosphere at a short distance from the sunspot. The other half of the structures are found to be ever rising in the atmosphere with increasing distance from the sunspot and therefore should form part of the super-penumbral canopy. All associated velocity fibrils seem to terminate in accelerating down-flow patches, regardless of their connectivity at the end farthest from the sunspot, with a bright head at the inner end. For a structure without a clear signature in the velocity, we determined that it consists of two short loops in temperature that directly connect two opposite-polarity patches outside of the penumbra in the so-called sunspot moat.
\section{Discussion}\label{secdiss}
\subsection{Applicability of analysis approach}
Determining the 3D topology in the solar atmosphere requires the use of spectral information from different formation heights in order to acquire the height stratification of solar properties. Our approach of analysing \ion{Ca}{ii} IR spectra is limited somewhat because of the LTE assumption, but seems satisfactory for retrieval of the topological structure. The exact temperature values and geometric heights (or optical depths) shown in the 3D cubes or 2D cross-sections are not particularly reliable at present. However, our finding that about 50\,\% of the features we examined in the super-penumbral canopy form short, arched loops should not be affected by our analysis approach. 

In region 1, the inversion shows a thermal structure composed of two separate loops of different length that lie low in the atmosphere and that connect down to the photosphere at both ends. This is corroborated by the photospheric magnetic field that has two patches of opposite polarity at the respective ends of the loops. In addition, the integration of the LOS field inclination predicts a shape of the magnetic field lines similar to the loops seen in the thermal structure.

The results of the LTE inversion allow us to trace the thermal connectivity of features and thus to distinguish between features that return to the photosphere and those that rise upwards out of the formation height range of Ca 854.2\,nm (cf.~Fig.~\ref{fig_cuts}).
\subsection{Implications for super-penumbral structure} 
\paragraph{Mass flows} The relatively high fraction of 50\,\% or more of individual fibrils that connect back down to the photosphere in a short distance from the sunspot  helps to explain the mass balance of the chromospheric inverse Evershed flow. If the majority of the field lines would extend high up into the atmosphere, it would be difficult to maintain any substantial mass inflow over long periods because of the steep density gradient in the solar atmosphere. Field lines that connect back down to the dense photosphere in the close proximity of the sunspot, where additionally the magnetic field strength at the inner foot point inside the penumbra is presumably larger than at the outer foot point, fulfill all necessary conditions for the development of siphon flows \citep[][]{cargill+priest1980,noci1981,thomas1988,degenhardt1989,doyle+etal2006}. These flows suck up photospheric material at the outer foot point and transport it into the sunspot. 

The LOS velocity at the inner down-flow patches is about 4\,km\,s$^{-1}$, well below the sound speed (Figs.~\ref{fig1} and \ref{fig7}).  Most of the loops that we examined were rather short (a few Mm only) and lie low in the atmosphere, thus the height difference between the foot points and the apex might not be sufficient to generate the supersonic flows observed for larger-scale siphon flows \citep{uitenbroek+etal2006,beck+etal2010a,bethge+etal2012}. The increase of the red-shift at the inner foot point of the fibrils in the penumbra need not indicate an acceleration, but could only reflect the orientation of the flow direction relative to the LOS. An exact calculation of the de-projected flow speed is beyond the scope of this paper, but would be possible assuming field-aligned flows and using the inversion results of the photospheric polarimetric data.
\paragraph{Thermal topology of fibrils}
While the flow pattern can suffer from LOS effects which vary between the limb and the center side of the sunspot, the thermal structure of fibrils in all directions around the sunspot is similar. They terminate in a local brightening whose height in the atmosphere reduces closer to the sunspot \citep[see also][]{louis+etal2014}. Some of the fibrils show a central, slightly darker lane flanked by lateral brightenings. However, this darkening is not seen for all fibrils in the present observations, perhaps because of an overlap of individual fibrils. More isolated examples of such chromospheric velocity and intensity fibrils in Ca 854.2\,nm \citep[e.g.][]{beck+etal2010a} show the typical lateral structuring more clearly. The general appearance of these chromospheric fibrils makes them similar to the dark-cored filaments found in the photosphere in the penumbra of sunspots \citep{scharmer+etal2002,suetterlin+etal2004,langhans+etal2007,bellotrubio+etal2007}. The common explanation for the appearance of dark cores in penumbral filaments \citep{scharmer+etal2008}, umbral light bridges \citep{berger+berdyugina2003,lites+etal2004} or umbral dots \citep{bharti+etal2007,sobotka+puschmann2009,louis+etal2012} is a density difference between the central axis and the sides caused by a convective upwelling in the middle of the structure with lateral down-flows. For dark-cored chromospheric fibrils this explanation, however, is not applicable because the atmosphere is optically thin outside of the line cores of strong chromospheric lines at the height at which these structures are found. In particular, no traces of lateral down-flows are found, and a central convective upwelling is lacking because the underlying regular convection shows no relation at all to the fibrillar structure \citep[cf.~Fig.~\ref{fig7} or][]{beck+etal2010a}. 

A second possibility for the creation of a density and/or opacity difference in such structures was given by \citet{ruizcobo+bellotrubio2008}. They show that in a hydro-dynamic model of a flux tube in the penumbra the flow speed along the structure can cause a similar opacity effect. Given that the LOS velocity clearly indicates a significant mass flow along the fibrils, this explanation for the appearance of dark cores seems more likely for chromospheric fibrils, with the caveats that in this case we observe an inflow into the sunspot, i.e.~an opposite direction of the flow relative to \citet{ruizcobo+bellotrubio2008}, and we must also explain the bright end point of the fibril. A possible explanation for the latter is a standing shock front where the chromospheric material impacts on dense photospheric material \citep{thomas+montesinos1991,degenhardt+etal1993,uitenbroek+etal2006,lagg+etal2007,bethge+etal2012}, which can best be traced with more suitable line blends to follow the velocity with height \citep{bethge+etal2012}. 
\subsection{Regions 1 and 4: quantitative information on topology}
For the structures located in regions 1 and 4, the LTE inversion retrieves some quantitative information on their topology. Region 4 is centred on a photospheric network element (Figs.~\ref{fig1} and \ref{fig5}). Theoretical models of magnetic flux concentrations \citep[e.g.][]{solanki+etal1991} predict a lateral expansion with increasing height in the atmosphere. Only recently was this confirmed directly from both observations and simulations \citep{beck+etal2013a,beck+etal2013c,delacruzrodriguez+etal2013}. In this context, the snapshot of the thermal topology of the network element in Fig.~\ref{fig16} is actually less interesting than the prospect of tracing such structures with time using imaging spectroscopic data which provide temporal resolution in a 2D FOV. 

Region 1 contains a structure that differs from the typical fibrils by its lack of systematic and/or fast flows. The LTE inversion, and also the appearance in the spectra of a weak signature in the line-core image and a strong signature at wavelengths slightly towards the line wing, suggest some low-lying feature with exceptional high temperature. Given the location outside the sunspot, the feature resembles event No.~5 in \citet{reardon+etal2013}, comparing for instance the line-core image of Ca 854.2\,nm in their Fig.~12 with the corresponding panel for region 1 in Fig.~\ref{fig7}. The LTE inversion gives a quantitative view of the thermal topology of the structure that consists of two neighbouring, short loops (Figs.~\ref{fig9} and Fig.~\ref{fig13}) which connect two magnetic patches of opposite polarity outside of the spot. It leaves, however, a few important questions open: why is the structure bright ? How is it evolving, is it rising or submerging ? Again, the slit-spectrograph data suffers here from a lack of temporal resolution which could provide more details on the exact physical process that occurred. 

An application of the LTE inversion, or any possibly improved non-LTE approach, to time-series of Ca 854.2\,nm spectra taken with imaging spectrometers such as IBIS, CRISP or the GFPI might therefore be interesting. For structures such as those found in region 1, their temporal evolution is a critical ingredient to determine the underlying physical process or driver. The structure could represent a buoyant rise of small-scale loops with a subsequent reconnection with overlying magnetic field that heats up the entire loop length, but from the current set of data such a behavior cannot be determined.
\section{Conclusions}\label{secconc}
We provide evidence that the inverse Evershed flow in the super-penumbral canopy at the formation height of Ca 854.2\,nm, i.e.~the low chromosphere as opposed to structures seen in the  H$\alpha$  line core, mainly reflects siphon flows along short loops that connect photospheric foot points. The dark-cored structure of the chromospheric fibrils cannot be caused by convective motions but should result from an opacity difference between the central axis and lateral parts of the fibril caused by the significant flow speed.
\begin{acknowledgements}
The Dunn Solar Telescope at Sacramento Peak/NM is operated by the National Solar Observatory (NSO). NSO is operated by the Association of Universities for Research in Astronomy (AURA), Inc.~under cooperative agreement with the National Science Foundation (NSF). HMI data are courtesy of NASA/SDO and the HMI science team. R.R.~acknowledges financial support by the DFG grant RE 3282/1-1. D.P.C.~acknowledges partial support by NSF grant ATM-0548260. We thank R.~Rutten for his critical review of the first draft of this article. We thank A.~Tritschler for providing us with the IDL routines for the calculation of the DST telescope polarization. We thank S.~Keil for his language editing efforts.
\end{acknowledgements}
\bibliographystyle{aa}
\bibliography{references_luis_mod}
\begin{appendix}
\section{LTE archive fit performance}\label{perform_fit}
\begin{figure*}
\resizebox{8.4cm}{!}{\includegraphics{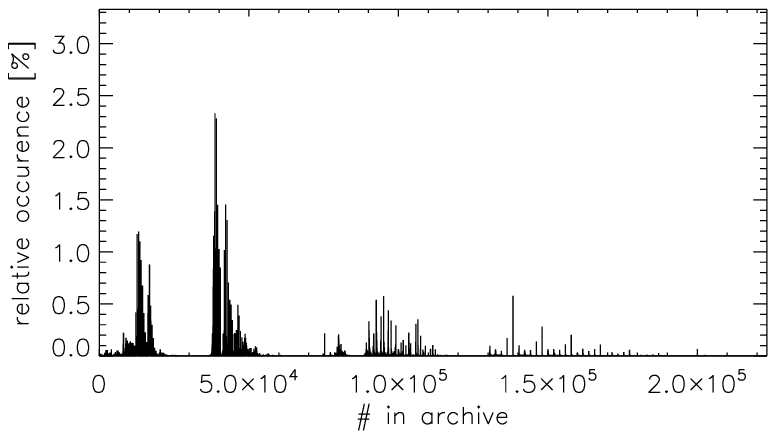}}\resizebox{8.4cm}{!}{\includegraphics{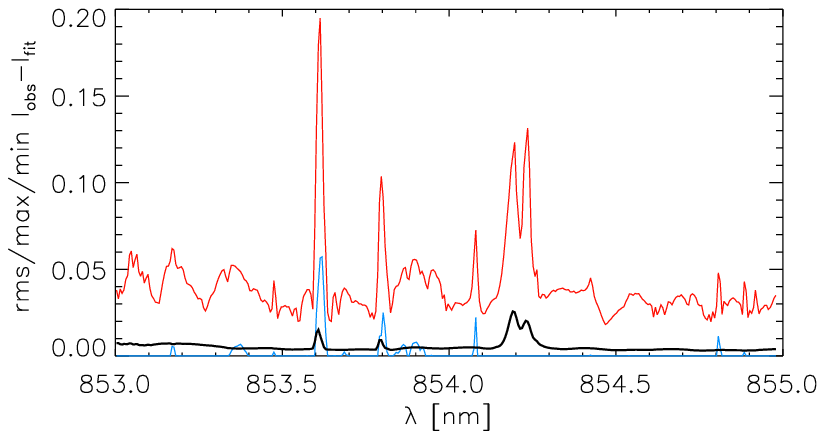}}
\caption{Left: Archive usage. Only 16\,705 out of 276\,802 profiles were used. Right: Maximal (red), minimal (blue), and rms deviation (thick black) between observed and best-fit intensities as function of wavelength.\label{arch_use_fig}}
\end{figure*}

In this section, we quantify the performance of the LTE archive fit to the observed spectra.
\paragraph{Archive usage} Similar to the result in BE13, only a small fraction of all archive profiles was used in the fit. From the 276\,802 profiles available, only 16\,705  ($\sim$6\,\% of the full archive) different ones were used in the least-square fit to the 167\,200 observed profiles. The left panel of Fig.~\ref{arch_use_fig} shows the distribution of the profiles used across the full archive. The profiles used cluster in ranges related to the generation of the archive's temperature stratifications from an initial model with an additional Gaussian perturbation (cf.~BE13). The lack of low-temperature profiles in the archive produces a mismatch between observed and best-fit profiles in parts of the umbra. Thus there is a need to include more low-temperature profiles in the archive to successfully fit profiles from active regions with sunspots and pores if no subsequent iterative modification of the temperature stratification as in BE13 is applied.
\paragraph{Fit quality}
\begin{figure*}
\centerline{\resizebox{11.8cm}{!}{\includegraphics{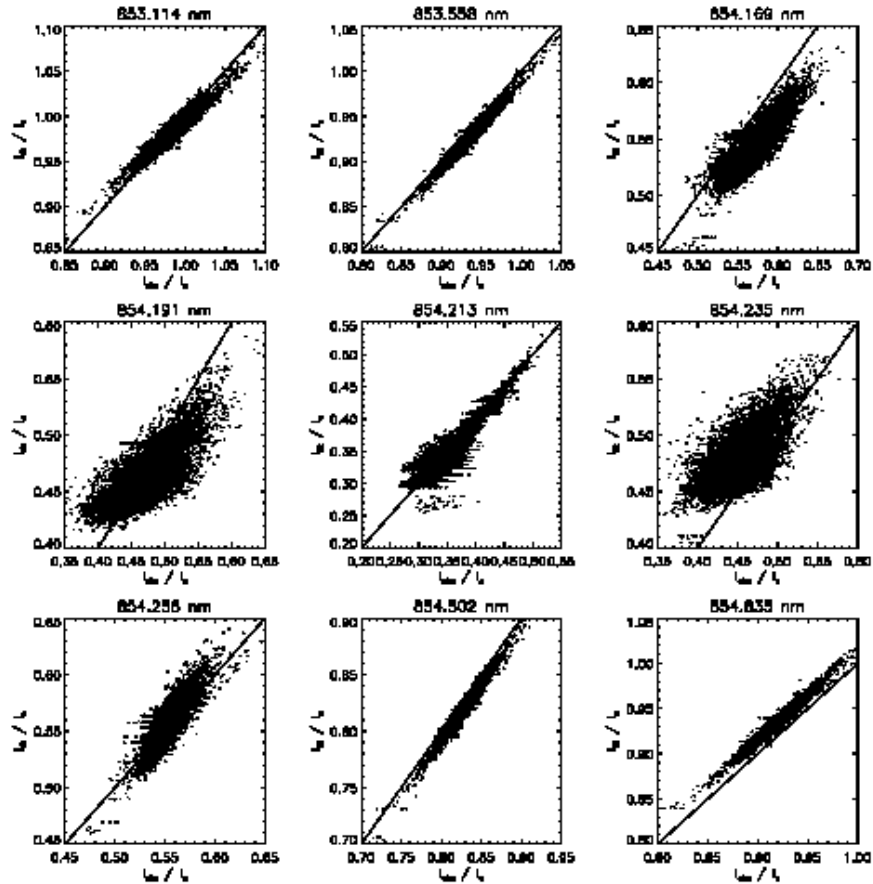}}}
\caption{Scatter plot of observed and LTE archive best-fit intensities at nine different wavelengths. The black line indicates the unity relation. \label{fit_scatter}}
\end{figure*}
One way to quantify the quality of the fit using the LTE archive is a calculation of minimal, maximal, and rms deviation between observed and best-fit intensities (right panel of Fig.~\ref{arch_use_fig}). As already seen by comparing Figs.~\ref{fig5} and Figs.~\ref{fig6}, the deviation depends on the wavelength, i.e.~close to the line core of any spectral line the deviations are larger than for blend-free wavelength windows in the line wing. On average, the rms deviation between observed and best-fit profiles was about 0.5\,\% of $I_c$, with maximal deviations of up to $> 10\,$\% in the line core of \ion{Ca}{ii} IR at 854.2\,nm, where the rms deviation was about 3\,\% of $I_c$.

The scatter plots of observed and best-fit intensities (Fig.~\ref{fit_scatter}) show the same trend, with a close fit for pseudo-continuum wavelengths in the line wing, and increasingly larger deviations near the line core. In the very line core at 854.213\,nm (central panel of Fig.~\ref{fit_scatter}), the lack of low-intensity profiles as well as the discrete nature of the archive is apparent in the form of horizontal stripes with gaps in between for $I < 0.32\,I_c$. Overall, the quality is still satisfactory, given the limitations of the approach.
\begin{figure*}
\centerline{\resizebox{17.6cm}{!}{\hspace*{1cm}\includegraphics{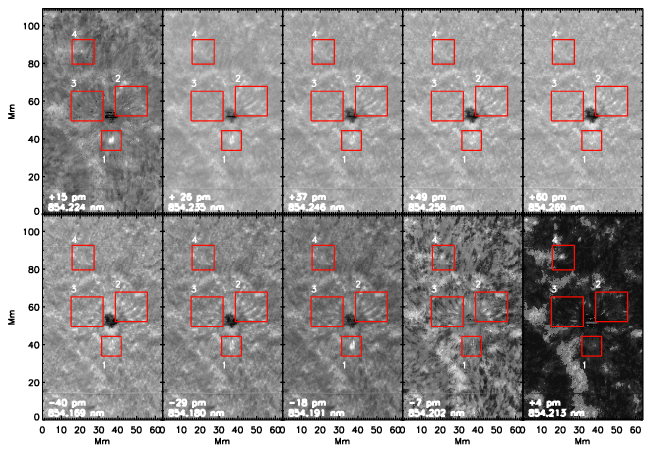}}}$ $\\
\caption{Same as Fig.~\ref{fig5} for the archive best-fit spectra with the SIR LOS velocities included.\label{fig6}}
\end{figure*}
A visual comparison of the quality of the fits and the effect on spatial patterns is provided by comparing Figs.~\ref{fig5} and \ref{fig6}. Figure \ref{fig5} shows the FOV in various wavelengths close to the line core of Ca 854.2\,nm in the observed spectra, while {Fig.~\ref{fig6}} shows the same in the best-fit archive spectra, after including the LOS velocities obtained from the SIR inversion in the synthesis. Analogously to the animation of temperature above $\log\,\tau <-4$, the best-fit spectra show a coarser spatial structuring near the line core where the resolution of the archive, in terms of the included number of different temperature stratifications, is insufficient to cover all of the observed spectra.
\section{3D rendering of network element}
\label{3dnetwork}
\begin{figure*}
\centerline{\hspace*{0cm}\resizebox{3.2cm}{!}{\includegraphics{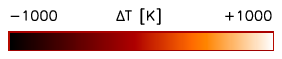}}}
\centerline{\resizebox{5cm}{!}{\includegraphics{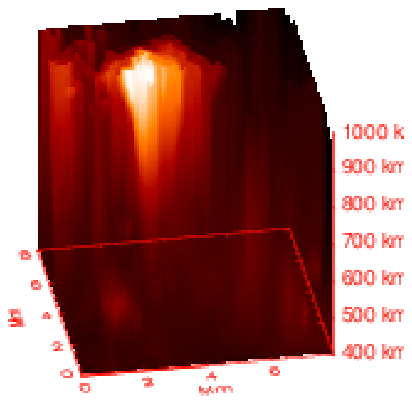}}\hspace*{.5cm}\resizebox{5cm}{!}{\includegraphics{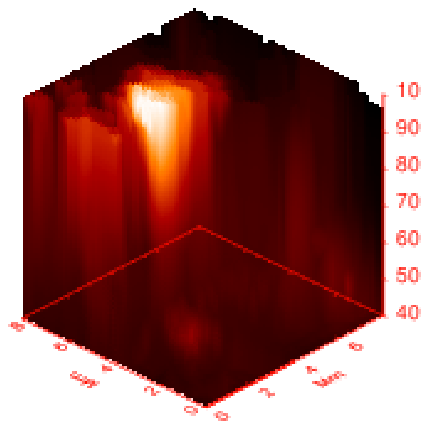}}\hspace*{.5cm}\resizebox{5cm}{!}{\includegraphics{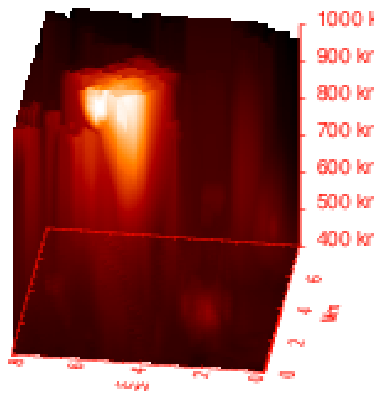}}}
\centerline{\resizebox{5cm}{!}{\includegraphics{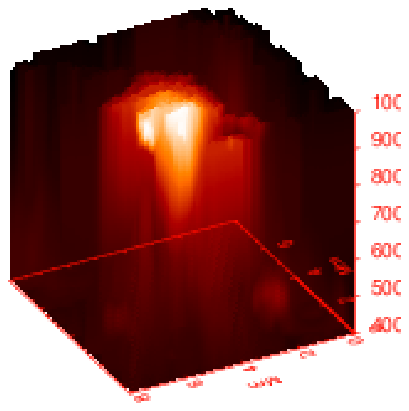}}\hspace*{.5cm}\resizebox{5cm}{!}{\includegraphics{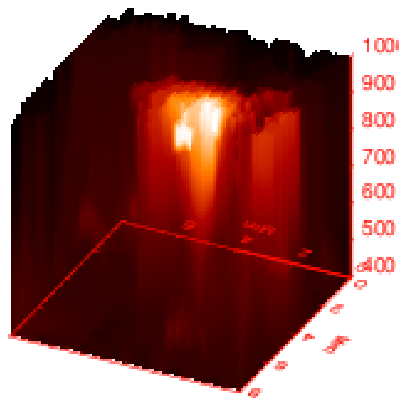}}\hspace*{.5cm}\resizebox{5cm}{!}{\includegraphics{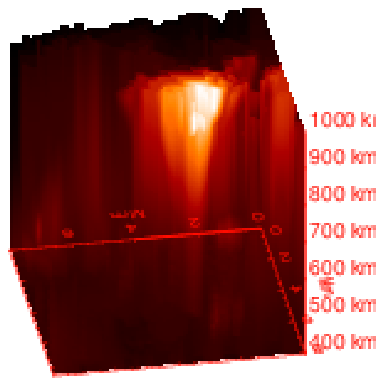}}}
\centerline{\resizebox{5cm}{!}{\includegraphics{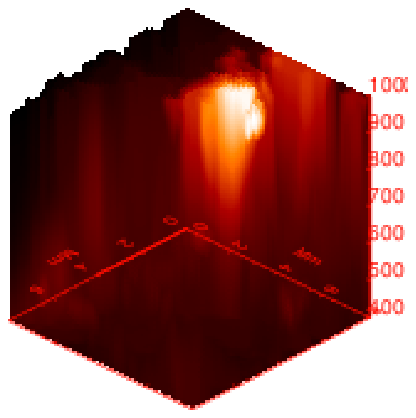}}\hspace*{.5cm}\resizebox{5cm}{!}{\includegraphics{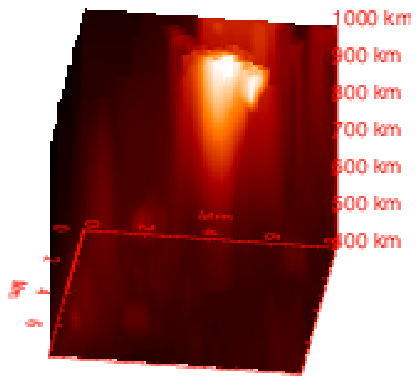}}\hspace*{.5cm}\resizebox{5cm}{!}{\includegraphics{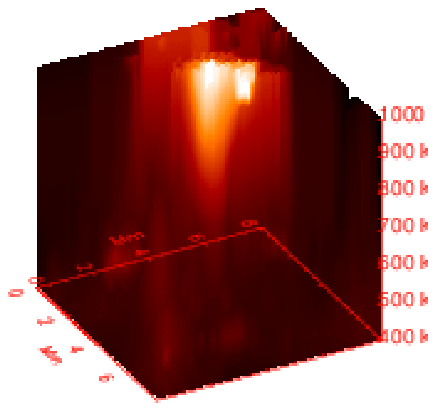}}}
\caption{3D rendering of the temperature difference from the average temperature stratification in region 4 from different viewing angles. Viewing angle relative to the horizontal plane is 35 deg. The angle relative to the $z$-axis increases by 36 deg from left to right in each row, and from top down. Display range is $\pm$\,1000\,K.\label{fig16}}
\end{figure*}

This section describes the derived thermal 3D topology of a roughly isolated network element. Figure \ref{fig16} does not show the modulus of temperature in the 3D rendering as in Figs.~\ref{fig8} to \ref{fig14}, but instead the difference of the individual temperature stratifications from the temperature stratification averaged over the full FOV. In the modulus of temperature, the structure of the network concentration is less apparent because it is not much hotter than the surroundings, especially when the LOS reaches down to the much hotter photosphere. Additionally, Fig.~\ref{fig16} displays more viewing angles than in the previous figures, going completely around the structure to enhance the visibility of the 3D topology. In the temperature difference cubes of Fig.~\ref{fig16}, the lateral expansion of the structure with height can be clearly seen, as well as other temperature enhancements behind the brightest one (cf.~Fig.~\ref{fig5}; region 4 is centred on the last brightening in a chain of photospheric magnetic flux concentrations). Note that the thermal canopies around network elements are actually more extended than it appears to be the case here \citep{beck+etal2013a,beck+etal2013c,delacruzrodriguez+etal2013}. This is because in the 3D rendering the highest value encountered along each LOS is used to define the appearance in the 3D cube making only the central part of the network element stand out. The smaller lateral expansion in the 3D rendering compared to 2D $x-z-$plots of temperature suggests that the central axis of the flux concentration is hotter than its canopy, i.e.~a radial temperature gradient should be present inside of the structure.
\end{appendix}
\end{document}